\documentclass[
reprint,
superscriptaddress,
longbibliography,
showkeys,
 amsmath,amssymb,
 aps,
pre, 
]{revtex4-2}

\usepackage{graphicx}
\usepackage[caption=false]{subfig} 

\usepackage{float}
     
\usepackage{dcolumn}
\usepackage{bm}

\newcommand{\ee}{\end{equation}} 
\newcommand{\be}{\begin{equation}}

\usepackage[mathscr]{euscript}  
\usepackage[dvipsnames]{xcolor}

\usepackage{tcolorbox}
\usepackage{comment}
\usepackage{float}

\makeatletter
\newsavebox{\@brx}
\newcommand{\llangle}[1][]{\savebox{\@brx}{\(\m@th{#1\langle}\)}%
  \mathopen{\copy\@brx\kern-0.5\wd\@brx\usebox{\@brx}}}
\newcommand{\rrangle}[1][]{\savebox{\@brx}{\(\m@th{#1\rangle}\)}%
  \mathclose{\copy\@brx\kern-0.5\wd\@brx\usebox{\@brx}}}
\makeatother

\begin{document} 

\preprint{APS/123-QED}

\title{Information bound on navigation speed in smart active matter}

\author{Kristian St\o{}levik Olsen}
\thanks{\textit{Correspondence: kristian.olsen@hhu.de}}
\affiliation{Institute for Theoretical Physics II - Soft Matter, Heinrich Heine University Düsseldorf, D-40225 Düsseldorf, Germany}

\author{Mitsusuke Tarama}
\affiliation{Department of Physics, Kyushu University, Fukuoka 819-0395, Japan}

\author{Hartmut L\"{o}wen}
\affiliation{Institute for Theoretical Physics II - Soft Matter, Heinrich Heine University Düsseldorf, D-40225 Düsseldorf, Germany}

\begin{abstract}
Intelligent behavior in life‑like systems often arises from the ability to gather, process, and act on information. While active matter provides a framework for studying life-like dynamics, it typically omits internal information‑processing and decision‑making. Here we introduce an adaptive active particle model that uses minimal information processing capabilities in order to navigate towards a distant target. By combining renewal‑based intermittent motion with the Cram\'{e}r-Rao inequality, we derive a bound on the navigation speed valid for a wide range of information processing strategies. The framework captures hallmark features of cognitive systems, including optimal sensing durations and a speed-accuracy trade-off that balances noise and reliability. Allowing stored information to degrade before action reveals that although deterioration slows navigation, the trade‑off remains governed primarily by external orientational noise and is remarkably insensitive to memory decay.
\end{abstract}

\maketitle

\textit{Introduction} --- Living and life-like systems are remarkable in their ability to perform purposeful behavior in complex or noisy environments. Such behavior emerges from the ability to sense, process, and act upon information about the environment, a crucial property not shared by passive counterparts \cite{schrodinger2025life}. Understanding and characterizing how such behavior emerges from the interplay between information and dynamics is central to ongoing efforts to extend the field of active matter into the realm of intelligent systems \cite{teVrugt2026artificial,baulin2025intelligent,kaspar2021rise}.

Active matter provides a powerful framework within statistical physics for modeling the dynamics of self-driven biological agents, and is based on the ability to convert energy input into sustained motion \cite{ramaswamy2010mechanics}. While such energy processing is a crucial property of life-like systems, so is the ability to process information \cite{tkavcik2016information}. Whether in molecular motors adjusting their steps to environmental cues \cite{toyabe2011thermodynamic,ariga2021noise} or in large animals foraging across complex landscapes \cite{viswanathan2011physics,volpe2017topography}, information underpins the physical behavior and efficiency of these systems.

However, only recently has the field begun to grapple with the informational aspects of behavior. Early developments such as chemotaxis-inspired models laid the groundwork for incorporating goal-oriented navigation into particle dynamics \cite{berg1977physics,schnitzer1993theory,de2004chemotaxis}.
Recent approaches have explored extensions of active matter by endowing particles with memory, internal states, or adaptive rules, leading to models often referred to as smart or intelligent active matter \cite{tsang2020roads,goh2022noisy,negi2025binary,rode2024information,cocconi2025dissipation,hou2025ornstein,sinha2025optimal,mori2025optimal,del2025proxitaxis,song2026ornsteinuhlenbeckinformationparticlenew}. Machine and reinforcement learning methods have also been used to reveal optimal strategies for navigating complex environments in scenarios where analytical treatment is intractable \cite{colabrese2017flow,jacob2025mixing,gassner2023noisy,putzke2023optimal,cichos2020machine,gustavsson2017finding,schneider2019optimal,durve2020learning,nasiri2023optimal,nasiri2024smart,heinonen2025optimal,singh2026homingreinforcementlearning}. These efforts aim to bridge the gap between passive response and active decision-making. Beyond theoretical insights, they are also relevant to real-world applications such as in autonomous robotics, where processing partial or noisy information in real time to navigate toward a goal is key, as well as to biological systems like insects 
that rely on incomplete internal representations to adapt their dynamics \cite{muller1988path,hartmann1995ant,lenz2012spatiotemporal,wiltschko2023animal}.

In parallel to these developments, several disciplines have developed formal frameworks to understand decision-making under uncertainty. In quantitative neuroscience and behavioral science, minimal models capture generic features such as  the tendency of agents to balance speed and accuracy when making time-sensitive decisions \cite{garrett1922study,ratcliff2008diffusion,ratcliff2016diffusion,siggia2013decisions,chittka2003bees,durmaz2023human,brunton2013rats}. Furthermore, search and navigation strategies based on Bayesian inference have been used to model chemotaxis and other sensory-guided behaviors \cite{vergassola2007infotaxis,celani2014odor}. Yet despite their relevance, such concepts are rarely integrated directly into active matter models in an analytically tractable way, and consequently information is often implicit or absent altogether.

In this work, we propose a model that combines classical active matter models and information theory by allowing particles to perform simple inference of a target direction. Within this framework, we demonstrate the existence of an information-constrained bound on the effective navigation speed, originating from the Cram\'{e}r-Rao inequality \cite{cover1999elements,kullback1997information}, and use this to analyze different information-processing strategies. Generic behavioral features such as the speed-accuracy trade-off, reflecting the tension between acting quickly and acting reliably, and the existence of optimal information processing times emerge from the model, demonstrating how minimal inference properties combined with active dynamics can account for generic behavioral features. Despite the simplicity of our model, it reveals rich behaviors, and offers a tractable setting for studying how information shapes active dynamics.\\

\begin{figure}[t!]
    \centering
    \includegraphics[width = 0.98\columnwidth]{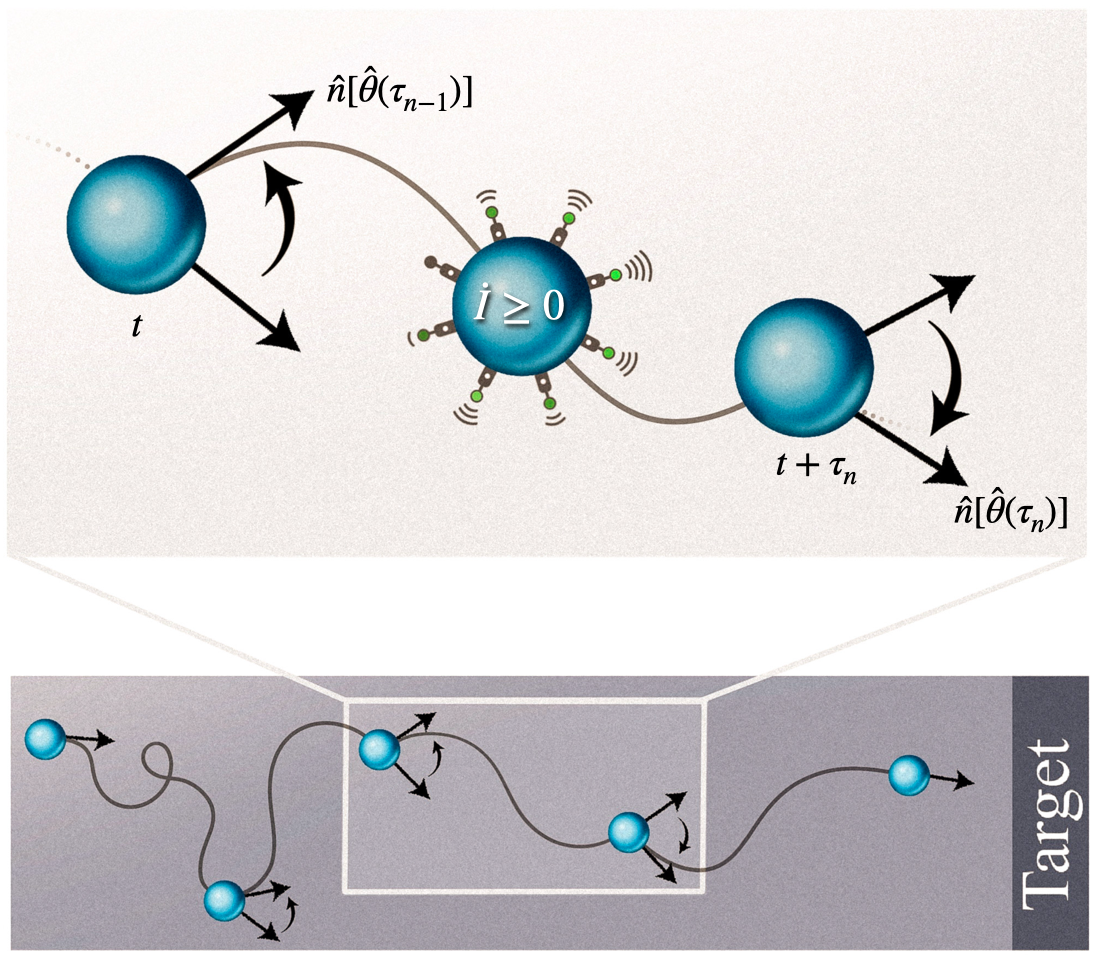}
    \caption{Active agents use an intermittent navigation strategy that relies on information acquired during statistically independent periods to episodically reorient, generating a drift in a target direction. The $n$'th independent epoch is characterized by an initial orientation $\hat n [\hat \theta(\tau_{n-1})]$ based on information from the previous epoch, after which the particle performs normal active Brownian motion for a time $\tau_n$ while collecting information at a rate $\dot I\geq 0$. }
    \label{fig:fig1}
\end{figure}

\textit{Information-based navigation strategies and Cram\'{e}r-Rao speed limits } ---   We consider a renewal-based myopic navigation strategy, namely one where the searcher has no long-term memory of the target direction. Rather, it collects information for a duration $\tau$, drawn from a \emph{renewal time} distribution $\psi(\tau)$, then proceeds by acting on this information and effectively erasing it. This scheme endows the system with a renewal structure that, while sufficient to incorporate and produce realistic effects, also allows analytical analysis. Dynamically, the strategy is biphasic and consists of two repeated phases; (1) runs with information acquisition and (2) informed reorientations (see Fig.\ \ref{fig:fig1}):
\begin{enumerate}
    \item \textit{Runs \& information acquisition:} During the $n$'th run, which last for a random time $\tau_n$ drawn from the renewal time density $\psi(\tau_n)$, the particle follows the dynamics of an active Brownian particle in two dimensions \cite{howse2007self,te2025metareview}:
\begin{align}
    \dot {\vec{x}}(t) &= v_0 \hat n(t),\label{eq:s1}\\
    \dot \phi(t) &= \sqrt{2 D_r}\xi(t),\label{eq:s2}
\end{align}
where $v_0$ is the self-propulsion speed, and $\hat n(t) = [\cos \phi(t),\sin \phi(t)]$ is the unit orientation vector pointing in the particle's direction of motion $\phi(t)$. The angular dynamics in the run phase is simply determined by a Gaussian white noise $\xi(t)$ with rotational diffusion coefficient $D_r$. We assume the particle to be capable of performing noisy measurements of the true target direction throughout the duration of this phase. The initial particle orientation in the $n$'th phase is an estimate of the true target direction based on the previous run phase, which we will denote $\hat\theta (\tau_{n-1})$.

\item \textit{Inference \& reorientations:} During the $n$'th run, the particle has acquired information about the true target direction based on noisy measurements or monitoring. Once this run phase is completed, the particle uses the collected information and reorients in a direction towards the target, which becomes the initial orientation for the $(n+1)$'th phase. Concretely, if a set of noisy measurements $\{\theta_j\}$ was made in time $\tau_n$, an estimator $\hat \theta(\tau_n)$ is a function of the measurements that estimates the true target direction $\theta$. The precision of any unbiased estimator ($\langle \hat \theta (\tau_n)\rangle = \theta$) is fundamentally limited through the Cram\'{e}r-Rao inequality  $\text{Var}(\hat{\theta}(\tau_{n}) )\geq I^{-1}(\tau_{n})$, relating precision to the reciprocal of the Fisher information. For a single measurement $\theta_j$ of Gaussian data with variance $\sigma_j^2$, one has simply $I_j = \sigma_j^{-2}$, and the total Fisher information is additive under independent measurements so that with $n_\tau$ measurements in time $\tau$ we have $I(\tau) = \sum_{j=1}^{n_\tau}\sigma_j^{-2}$\cite{cover1999elements,kullback1997information}.
\end{enumerate}

Several natural systems move akin to the above biphasic rules, for example insects using the combination of wind direction and odors to orient \cite{carde2008navigational}, or dung beetles that intermittently collect information about new directions of motion from the environment or even the milky way \cite{byrne2003visual,dacke2013dung}. The renewal structure of this model, i.e., the independence of each epoch, enables late-time dynamical properties to be obtainable from single-epoch properties, a property recently utilized in recent works to study rectified motion in the absence of information processing \cite{kumar2020active,baouche2024active,olsen2024optimal,shee2025steeringchiralactivebrownian,kundu2025emulating}. If the particle covers a mean distance $\ell_\parallel$ in the direction of the target in a single run phase, the effective drift speed in the target direction can be measured by $v_\parallel = \ell_\parallel/\langle \tau \rangle$, and in dimensionless form $u_\parallel = v_\parallel/v_0$. Without loss of generality, we place the target direction parallel to the $x$-axis (see Fig.\ref{fig:fig1}). By combining the active Brownian dynamics with the Cram\'{e}r-Rao inequality, one arrives at a speed limit that must be obeyed for any renewal and information processing strategy:

\begin{equation}\label{eq:main}
    u_\parallel \leq u_\text{CR} = \frac{1-\tilde \psi(D_r)}{D_r \langle \tau \rangle} \int_0^\infty d\tau  \psi(\tau) \exp\left(-\frac{1}{2 I(\tau)}\right),
\end{equation}
This constitutes the central result of our work, and demonstrates general relations between particle persistence $D_r^{-1}$, renewal strategy (through $\psi(\tau)$ and its Laplace transform $\tilde \psi(s)$), information acquisition $I(\tau)$, and a largest allowed speed $u_\text{CR} = u_\text{CR}[\psi,I]$. The speed limit depends functionally on the tuple $\{\psi,I\}$, which determines the overall strategy of the active agent. By varying these a wide range of bounds can be derived, and generic well-established behavioral phenomena recovered, as discussed in the following. We note that the Cram\'{e}r-Rao bound is saturated for so-called efficient estimators, which means that we can also interpret the bound as the speed curves for particles following the most efficient strategy. For a discussion of these matters, as well as a derivation of Eq. (\ref{eq:main}), see the appendix.

In the following, we demonstrate the results through two natural applications, namely I) linear information growth with constant information acquisition rate $\dot I$, and II) information deterioration during internal storage. Within each information acquisition strategy, different choices of renewal durations can be explored by varying $\psi(\tau)$, revealing how particular strategies depend on stochasticity of the renewal durations.

\begin{figure}
    \centering
    \includegraphics[width=\columnwidth]{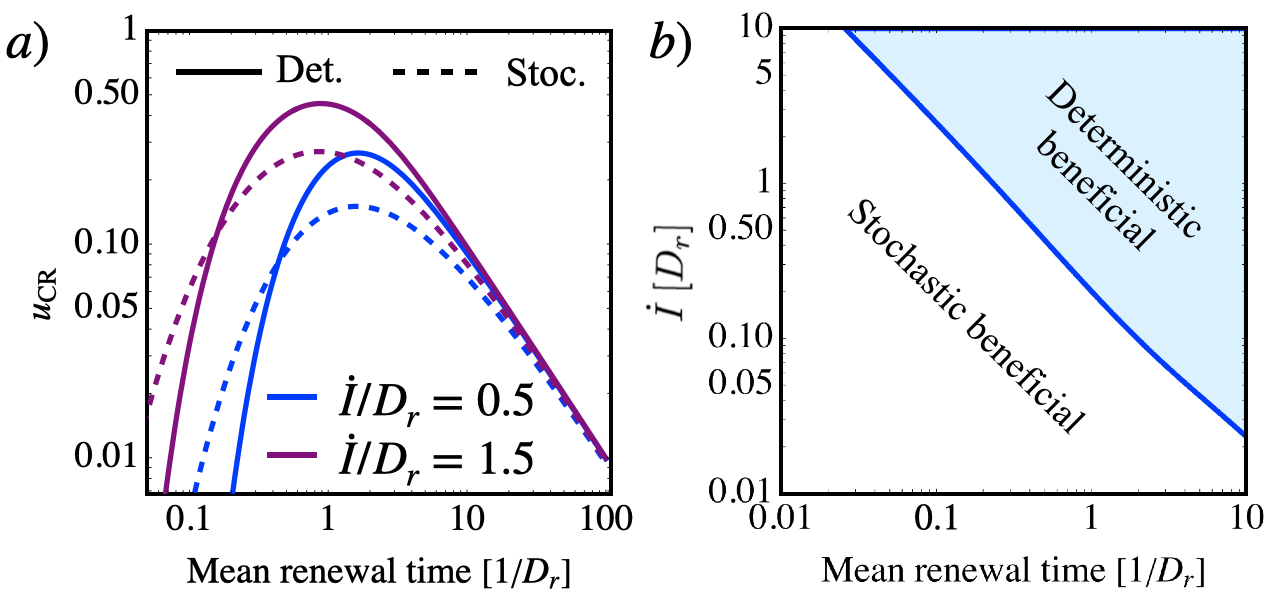}
    \caption{a) Cram\'{e}r-Rao bound for the navigation speed for two different values of the information per persistence time. Solid line shows a deterministic renewal strategy, while the dashed line a stochastic one. b) Phase diagram where deterministic vs. stochastic strategies maximize the possible navigation speed, as function of mean renewal time and information gained.   }
    \label{fig:detvsstoc}
\end{figure}

\textit{Case I: constant information rate} ---  Consider the case $I(\tau) = \dot I \tau$, representing a situation where the particle at short evenly distributed time intervals during the duration $(0,\tau)$ performs uncorrelated and identically distributed measurements of the true target direction. A derivation of this is contained by taking an appropriate limit of \textit{Case II} discussed below. In the simplest case of deterministic periods $\tau$, the bounding speed in Eq. (\ref{eq:main}) takes the form
\begin{equation}
    u_\text{CR}^\text{det.} = \frac{1-e^{-D_r \tau}}{D_r  \tau }  \exp\left(-\frac{1}{2 \dot I \tau}\right).
\end{equation}
The presence of rotational diffusion $D_r>0$ makes the drift speed decrease in time, while the presence of a positive information acquisition rate $\dot I >0$ makes it grow. These competing effects lead to a non-monotonic bound $u_\text{CR}$ as a function  of the run phase durations $\tau$, as seen in Fig. \ref{fig:detvsstoc}a. To gain analytical insights into this non-monotonicity, we consider momentarily the regime of highly persistent active particles, where $D_r \tau \ll 1$. Expanding the first factor in the bound leads to $u_\text{CR} \approx (1-D_r\tau/2) \exp\left(-\frac{1}{2 \dot I \tau}\right)$, which has a maximum at the optimal renewal duration $\tau_\text{opt} =  \frac{1}{4 \dot I}( \sqrt{\frac{16 \dot{I}}{D_r} +1} -1 ) $. The optimal renewal duration decays monotonically with $\dot I$, and at large $\dot I$ scales as ${\dot I}^{-1/2}$. We also see that increased noise, as measured by the rotational diffusion coefficient, will decrease the optimal sensing time, reflecting the negative effect of longer periods where noise is allowed to act. Particles that are more efficient in acquiring information should act upon this information more rapidly, as this avoids the negative effects of rotational diffusion.  

A relevant question in renewal-based search strategies in general is whether stochasticity in the renewal periods can enhance search \cite{bhat2016stochastic,evans2025stochastic}. To understand the role played by fluctuations in the sensing period, we consider Poissonian renewal times, in which case we have an exponential distribution $\psi(\tau) = r e^{- r\tau}$ characterized by the rate $r$. Using integral representations of modified Bessel function of the second kind $K_1(z)$\cite{NIST:DLMF}, we find from Eq. (\ref{eq:main}) the speed limit 
\begin{align}
  u_\text{CR}^\text{Pois.} =   \frac{r}{r+D_r} \sqrt{\frac{2 r}{\dot I}} K_1\left(\sqrt{\frac{2 r}{\dot I}} \right).
\end{align}
As in the deterministic case, the drift speed crosses over from increasing at small rates, to decaying at large rates, implying the existence of an optimal value of the reorientation rate $r$. The speed limits based on Poissonian renewal times are shown in dashed lines in Fig. \ref{fig:detvsstoc}a. We see that in both the deterministic and stochastic case, a trade-off takes place between acting rapidly in order to avoid noise and having longer durations that are more accurate. Within behavioral science this is referred to as a speed-accuracy trade-off, and is a feature found in several living systems across scales \cite{garrett1922study,ratcliff2008diffusion,ratcliff2016diffusion,siggia2013decisions,chittka2003bees,durmaz2023human,brunton2013rats}. 

While the stochasticity of the renewal duration does not affect the presence of a speed-accuracy trade-off, it is interesting to observe that in certain regimes variability can either be helpful or detrimental. By considering the two cases of deterministic and Poissonian durations at the same mean $\tau \overset{!}{=} 1/r$, a fair comparison can be made where only the fluctuations in $\tau$ in the Poissonian case differs. Fig. \ref{fig:detvsstoc}b shows regimes where the deterministic strategy is most beneficial. Generally, for small (mean) sensing periods, the particle will in the deterministic case not have collected much information, and fluctuations could help in this regard. At large sensing periods, fluctuations of the exponential type tends to decrease the sensing time of parts of an ensemble of realizations, leading to inefficient strategies.

\textit{Case II: Deteriorating information} --- The framework derived can be used to study sensing strategies beyond those where information simply grows linearly. If the active agent collects data as before, the data must be saved internally until the renewal period ends and the time comes for the particle to act on the information. Assuming that the internal memory storage is imperfect, the data may be corrupted or deteriorate over time. This raises an intriguing question regarding the speed-accuracy trade-off, which as we have seen, balances effects originating in noise and reorientation accuracy. How does the introduction of internal noise that acts on data enter into this trade-off? 

 We consider a sequence of measurements $\theta_j$ taken at evenly spaced times $t_j$ within the run phase $(0,\tau)$. The measurements are recorded with error $\sigma^2$ at the moment they are taken, representing limits to the particles sensory equipment. The stored data gradually deteriorates before it is used for inference at the end of the run phase. This degradation is modeled as an internal diffusion process with diffusivity $D_d$, which we interpret as a deterioration rate. As a result, each data point accumulates an additional uncertainty that grows linearly with the time it spends in storage: beyond the intrinsic sensing noise $\sigma^2$, a measurement taken at time $t_j$ acquires an extra variance proportional to the remaining time $\tau - t_j$. For convenience, we denote the incremental deterioration occurring between successive measurements by $\mathcal{D} = D_d \Delta t$. The inference problem therefore becomes one involving heteroskedasticity, since the total variance is no longer uniform across the data sequence at the time of inference. Focusing on the regime of very frequent sampling, where $\Delta t$ is small, one obtains the following expression for the Fisher information
\begin{equation}\label{eq:fish}
    I(\tau) = \frac{1}{2\mathcal{D}} \log(1 + 2\mathcal{D}\dot I \tau),
\end{equation}
\begin{figure}[t!]
    \centering
    \includegraphics[width=\columnwidth]{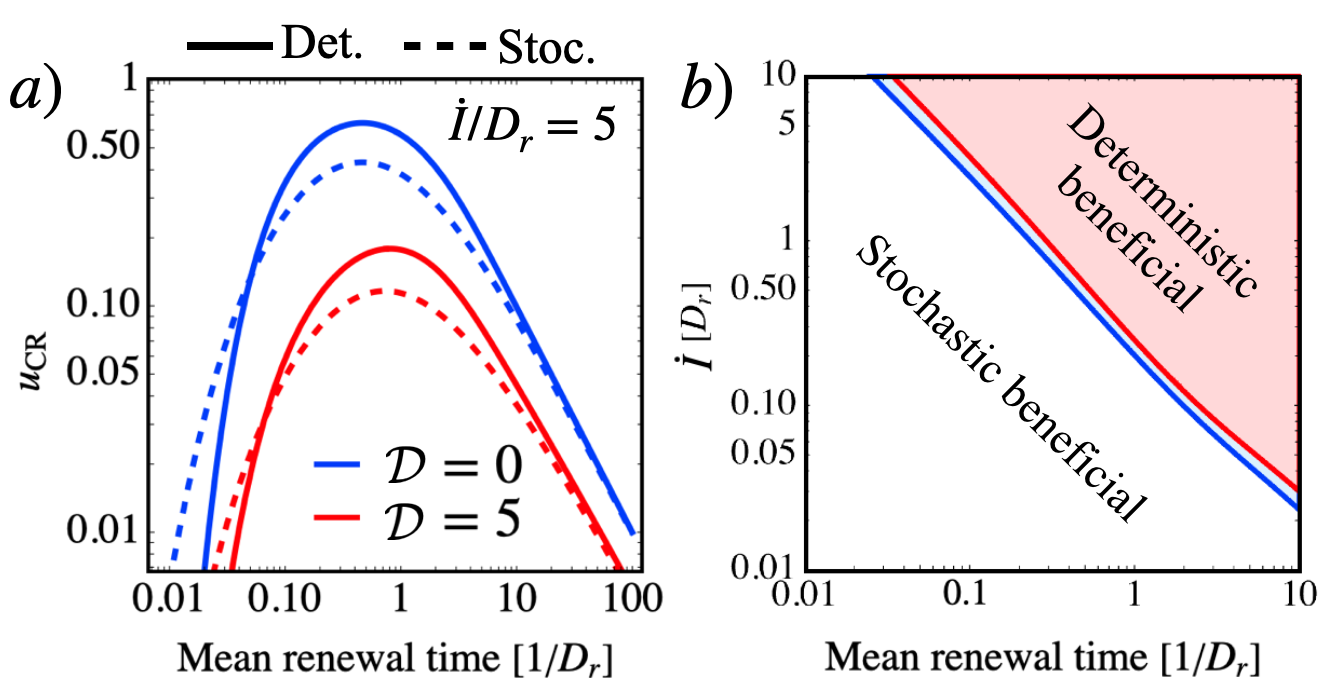}
    \caption{a) Cram\'{e}r-Rao bound for the navigation speed without ($\mathcal{D}=0$) and with $(\mathcal{D} = 5)$ deterioration of the data before inference. Solid lines show deterministic renewal strategies, while the dashed line corresponds to Poissonian renewal times. The dimensionless information rate is set to $\dot{I}/D_r = 5$.  b) Crossovers between optimal strategies. }
    \label{fig:deteriorate}
\end{figure}
where we introduced $\dot I = 1/(\sigma^2 \Delta t)$ as the information rate so that at early times $I(\tau) \approx \dot I \tau$.  This follows from the additivity of the Fisher information under independent measurements, and a careful derivation can be found in the appendix. We see that deterioration leads to ultra-slow logarithmic growth of the Fisher information as a function of $\tau$. The deterioration-less limit $\mathcal{D}\to 0$ recovers the previous case, which is readily verified by a simple Taylor expansion of the above logarithm. The slow logarithmic growth indicates that due to deterioration of the data, the active agent is subject to diminishing returns whereby longer $\tau$ gives new and accurate data-points but also lets old data become irrelevant.  In order for the particle to reach the bound set by the Cram\'{e}r-Rao inequality in this case, it must correctly weight the measurements when estimating the target direction. Given a discrete set of $n_\tau$ measurements $\{\theta_j\}$ in time $\tau$, the optimal estimator is given by $\hat \theta_\text{opt} = I^{-1}(\tau) \sum_{j=1}^{n_\tau} [\sigma^2 + 2D_d (\tau-t_j)]^{-1} \theta_j$. Hence to reach the bound the particle must not only know all the measurement times, but also the error of its sensory apparatus. This may make the bound harder to reach, although approaching the bound may be possible through evolutionary adaptation or other forms of learning. 

From our general results, Eq. (\ref{eq:main}), the above Fisher information leads to a speed limit which for deterministic sensing and run phases reads
\begin{equation}
    u_\text{CR}^\text{det.}(\mathcal{D}) = \frac{1-e^{-D_r \tau}}{ D_r \tau }  \exp\left(-\frac{\mathcal{D}}{ \log \left({1+2 \mathcal{D} \dot{I} \tau  }\right)}\right).
\end{equation}
Note that since $\log(1+x)\leq x$ for all positive $x$, we have $I(\tau) \leq \dot I \tau$. It hence follows that also $\exp(-1/(2I(\tau)))\leq \exp(-1/(2\dot I \tau))$, meaning that deterioration always suppresses the speed for all values of $\tau$. From this argument alone, deterioration may, however, still change the location of the maxima. We assume that, since large values of $\mathcal{D}$ would make any strategy slow, the realistic regime is that of relatively small values of $\mathcal{D}$. In this regime, we can expand $u_\text{CR}^\text{det.}(\mathcal{D})$ to leading order in $\mathcal{D}$ and arrive at $u_\text{CR}^\text{det.}(\mathcal{D}) \approx u_\text{CR}^\text{det.}(0)(1-\mathcal{D}/2)$. Since the leading order effect does not depend on $\tau$, the same holds for stochastic strategies $u_\text{CR}^\text{Pois.}(\mathcal{D}) \approx u_\text{CR}^\text{Pois.}(0)(1-\mathcal{D}/2)$. This shows that the leading order effect of deterioration is simply to re-scale the speed limit while leaving the optimal renewal duration $\tau_\text{opt}$ unchanged. In fact, Fig. \ref{fig:deteriorate}a shows that the optimal renewal duration, even in regimes of stronger deterioration, surprisingly does not change much. This demonstrates that deterioration of internally saved data through internal noise, while lowering the speed limit, does not significantly alter the speed-accuracy trade-off, which still seem to focus on a balance of \emph{external} noise and reorientation accuracy. Finally, as shown in Fig.\ \ref{fig:deteriorate}b, we see that the transition into the regime where deterministic strategies dominate also remains relatively untouched once deterioration is included, but is slightly shifted to larger values of the mean renewal time. In Fig.\ \ref{fig:deteriorate} we used Eq. (\ref{eq:main}) together with the Fisher information given by Eq.\ (\ref{eq:fish}) and numerically integrated over exponential renewal times to obtain the speed limit in the stochastic scenarios. This shows that while internal noise that deteriorates data can strongly suppress navigation speed, the optimal sensing duration that determines the speed-accuracy trade-off as well as the location of crossovers to regimes where fluctuating sensing periods may be beneficial remains sensitive mostly to the external dynamical noise.

\textit{Conclusions} --- We have analyzed a minimal model of active navigation that combines the renewal structure of intermittent active motion with information processing, revealing a speed-accuracy trade-off and optimal information acquisition strategies. We have derived an information-limited bound on the drift speed towards a stationary target direction, based on statistical inference performed by the particle. The bound, reachable through optimal strategies, is valid for a wide range of stochastic renewal durations and information processing strategies, and provides analytical insights into the combined effect of persistent active motion and information processing. 

Many future directions of research could be viable. Extensions that incorporate thermodynamics and enable trade-offs between accuracy, information as well as energy expenditure would be of interest \cite{nguyen2024remark,sartori2014thermodynamic,lan2012energy,welker2026accuracycomescostoptimal}. Here we have presented results based on the mean drift speed, and a better understanding of higher-order fluctuations would be insightful. For example, it may be of interest for the particle to have reliable low-fluctuating strategies even if this comes at the cost of reduced mean navigation speed, which raises interesting further optimization questions.

\acknowledgements
KSO is grateful to Benjamin Friedrich and Antonio Celani for insightful discussions. KSO acknowledges support from the Alexander von Humboldt Foundation. MT acknowledges financial support by JSPS KAKENHI Grant Number JP24H01485. HL acknowledges funding by the Deutsche Forschungsgemeinschaft (DFG) within the project LO 418/29-1. 

\appendix

\section{Derivation of the bound (Eq.\ref{eq:main})}
 From the active Brownian equations of motion, Eqs. (\ref{eq:s1}-\ref{eq:s2}), the  displacement during the $n$'th epoch is simply
\begin{equation}
    \Delta x_\parallel = v_0 \int_0^{\tau_n} dt\cos(\hat{\theta}(\tau_{n-1}) + \sqrt{2 D_r} W(t)).
\end{equation}
where $W(t)$ is a Wiener process with zero mean and variance $\langle W^2(t)\rangle = t$.
Performing the average over the Wiener process and using standard properties of trigonometric functions of Gaussian variables results in the conditional displacement
\begin{align}
   \langle \Delta x_\parallel| \hat{\theta}(\tau_{n-1}), \tau_n\rangle &= 
     v_0 \cos(\hat{\theta}(\tau_{n-1}) ) \frac{1-e^{-D_r\tau_n}}{D_r}.
\end{align}
Three random variables remain to be averaged in order to get the mean displacement; the two independent and identically distributed times $\{\tau_{n-1},\tau_n\}$, and the estimator $\hat \theta$. We first perform the average over the estimator, assuming that it is Gaussian distributed with zero mean. This assumes that, for example if measurements are Gaussian distributed, the particle acts on the collected information through some linear combination of the data. This includes strategies such as picking the median, mean, extrema or midpoint of the data to name a few. Averaging over the estimator gives a displacement depending only on the two statistically independent renewal times

\begin{align}\label{eq:temp1}
    \langle \Delta x_\parallel| \tau_{n-1}, \tau_n\rangle &= 
    v_0 \frac{1-e^{-D_r\tau_n}}{D_r} \exp\left(-\frac{\text{Var}(\hat{\theta}(\tau_{n-1}) ) }{2}\right).
\end{align}
The Cram\'{e}r-Rao bound from statistical inference tells us that the variance of this estimator is bounded by the reciprocal of the Fisher information $\text{Var}(\hat{\theta}(\tau_{n-1}) )\geq I^{-1}(\tau_{n-1})$. For a single measurement of data $\theta_j$ coming from a distribution $p(\theta_j;\theta)$ one has $I_j = \langle (\partial_\theta \log p(\theta_j;\theta))^2 \rangle$ \cite{cover1999elements,kullback1997information}. Rather than assuming an explicit model for the individual measurements made by the particle during the information acquisition phase, we will assume that the full phase leads to a total Fisher information $I(\tau)$ after many measurements. The functional dependence of $I(\tau)$ on $\tau$ defines the models information acquisition strategy. Since the exponential is monotonically decaying, the Cram\'{e}r-Rao bound can also be stated as $\exp\left(-\frac{\text{Var}(\hat{\theta}(\tau_{n-1}) ) }{2}\right) \leq \exp\left(-\frac{1 }{2 I(\tau_{n-1})}\right)$, directly offering a bound for the particle displacement in Eq. (\ref{eq:temp1}). Averaging over the two remaining independent and identically distributed durations to find the displacement $\ell_\parallel = \langle  \langle \Delta x_\parallel| \tau_{n-1}, \tau_n\rangle \rangle_{n-1,n}$ results in the speed-limit
\begin{equation}
    u_\parallel \leq u_\text{CR} = \frac{1-\tilde \psi(D_r)}{D_r \langle \tau \rangle} \int_0^\infty d\tau  \psi(\tau) \exp\left(-\frac{1}{2 I(\tau)}\right),
\end{equation}
 where we defined the (dimensionless) mean speed as $ u_\parallel =  v_\parallel/v_0 =  \ell_\parallel / (\langle \tau \rangle  v_0)$, and $\tilde\psi(s)$ is the Laplace transform of $\psi(\tau)$.

\section{Fisher information under internal memory deterioration}
Each data-point has, in addition to error $\sigma^2$ coming from the measurement itself, an additional variance $2D_d (\tau-t_k)$ due to the internal deterioration taking place before inference is made. Since the Fisher information is additive for independent measurements, it is given by the sum

\begin{align}
    I(\tau) &= \sum_{k=1}^n \frac{1}{\sigma^2 + 2D_d (\tau-t_k)} = \sum_{k=1}^n \frac{1}{\sigma^2 + 2\mathcal{D}(n-k)} ,
\end{align}
where we considered a discretization of size $\Delta t$ in time, so that $t_j =j \Delta t$, with $\tau = t_n = n \Delta t$, and used $\mathcal{D} = D_d \Delta t$. Note that by re-parametrizing the sum  in terms of $j = n-k$ we have
\begin{align}
    I(\tau) &= \frac{1}{2\mathcal{D}}\sum_{j=0}^{n-1} \frac{1}{\sigma^2/(2\mathcal{D}) + j} .
\end{align}
To proceed, we start by noting that repeated use of the recursive property $\Gamma(z+1) = z\Gamma(z)$  of the Gamma-function leads to the identity $\Gamma(z+n) = z^{(n)} \Gamma(z)$, where $z^{(n)}= z(z+1)\cdots (z+n-1)$ is the Pochhammer ascending factorial. Taking a logarithmic derivative gives
\begin{equation}
    \frac{d}{dz}\log\Gamma(z+n) = \sum_{j=0}^{n-1}\frac{1}{z+j} +  \frac{d}{dz}\log\Gamma(z).
\end{equation}
Re-arranging and applying this identity to our above sum results in
\begin{align}
    I(\tau)  &= \frac{1}{2 \mathcal{D}} \left [ \frac{\Gamma'(\frac{\sigma^2}{2\mathcal{D}}+n)}{\Gamma(\frac{\sigma^2}{2\mathcal{D}}+n)}- \frac{\Gamma'(\frac{\sigma^2}{2\mathcal{D}})}{\Gamma(\frac{\sigma^2}{2\mathcal{D}})} \right]. \label{eq:generalfisher}
\end{align}
We note that in the $\mathcal{D}\to \infty$ limit this reduces to $I(\tau) = 1/\sigma^2$, which is the Fisher information of a single Gaussian measurement at the final time $\tau$. In this limit, any measurement prior to the reorientation time at the end of the run phase immediately deteriorates and becomes irrelevant, and the particle only uses the most recent data point measured exactly at the time of reorientation. 

While Eq. (\ref{eq:generalfisher}) is exact and already can be used in Eq.\ (\ref{eq:main}) with $n = \tau/\Delta t$, we will for the sake of transparent interpretation simplify the expression with a minimal approximation. When the sampling rate is high $\Delta t \ll 1$ we can use the asymptotic expansion $\Gamma'(z)/\Gamma(z) = \log(z) + \mathcal{O}(z^{-1})$, leading to
\begin{equation}
    I(\tau) = \frac{1}{2\mathcal{D}} \log(1 + 2\mathcal{D}\dot I \tau),
\end{equation}
where $\dot I = (\sigma^2 \Delta t)^{-1}$, which is used and discussed in the main text.

\section{Maximal likelihood strategies}

The speed limit set by the Cram\'{e}r-Rao bound is tight, in the sense that known strategies can reach it. Indeed, for the case of independent and identically distributed measurements, the optimal estimates the agent can use is simply the arithmetic mean. For the case with deterioration, the optimal strategy is more complex, and depends on the time of the measurements. The optimal estimator is often the maximum likelihood estimator (MLE), obtained by solving $\partial_{\hat \theta_\text{MLE}} \log P(\{\theta_j\};\hat \theta_\text{MLE})=0$, where $P$ denotes the joint distribution of the data. Since each measurement is independent, $P(\{\theta_j\};\theta) = \prod_j p(\theta_j;\theta)$ with $ p(\theta_j;\theta)\sim \mathcal{N}(\theta,\sigma^2 + 2D_d(\tau-t_j))$. The MLE is therefore obtained by solving
\begin{equation}
    \sum_{j=1}^{n_\tau}\frac{\theta_j-\hat \theta_\text{MLE}}{\sigma^2+ 2D_d(\tau-t_j)} = 0.
\end{equation}
The solution is given by
\begin{equation}
    \hat \theta_\text{MLE} = \frac{1}{I(\tau)}\sum_{j=1}^{n_\tau}\frac{\theta_j}{\sigma^2 + 2D_d (\tau-t_j)} ,
\end{equation}
where $n_\tau$ is the number of measurements taken in the period $\tau$ and we recognized the Fisher information as the sum of inverse variances. From this one readily verifies that $\text{Var}(  \hat \theta_\text{MLE})  = 1/I(\tau)$. Hence the optimal strategy where $u_\parallel = u_\text{CR}$ must consider a weighted sum of the data, where old and more deteriorated measurements are given less weight.


\begin{thebibliography}{67}%
\makeatletter
\providecommand \@ifxundefined [1]{%
 \@ifx{#1\undefined}
}%
\providecommand \@ifnum [1]{%
 \ifnum #1\expandafter \@firstoftwo
 \else \expandafter \@secondoftwo
 \fi
}%
\providecommand \@ifx [1]{%
 \ifx #1\expandafter \@firstoftwo
 \else \expandafter \@secondoftwo
 \fi
}%
\providecommand \natexlab [1]{#1}%
\providecommand \enquote  [1]{``#1''}%
\providecommand \bibnamefont  [1]{#1}%
\providecommand \bibfnamefont [1]{#1}%
\providecommand \citenamefont [1]{#1}%
\providecommand \href@noop [0]{\@secondoftwo}%
\providecommand \href [0]{\begingroup \@sanitize@url \@href}%
\providecommand \@href[1]{\@@startlink{#1}\@@href}%
\providecommand \@@href[1]{\endgroup#1\@@endlink}%
\providecommand \@sanitize@url [0]{\catcode `\\12\catcode `\$12\catcode `\&12\catcode `\#12\catcode `\^12\catcode `\_12\catcode `\%12\relax}%
\providecommand \@@startlink[1]{}%
\providecommand \@@endlink[0]{}%
\providecommand \url  [0]{\begingroup\@sanitize@url \@url }%
\providecommand \@url [1]{\endgroup\@href {#1}{\urlprefix }}%
\providecommand \urlprefix  [0]{URL }%
\providecommand \Eprint [0]{\href }%
\providecommand \doibase [0]{http://dx.doi.org/}%
\providecommand \selectlanguage [0]{\@gobble}%
\providecommand \bibinfo  [0]{\@secondoftwo}%
\providecommand \bibfield  [0]{\@secondoftwo}%
\providecommand \translation [1]{[#1]}%
\providecommand \BibitemOpen [0]{}%
\providecommand \bibitemStop [0]{}%
\providecommand \bibitemNoStop [0]{.\EOS\space}%
\providecommand \EOS [0]{\spacefactor3000\relax}%
\providecommand \BibitemShut  [1]{\csname bibitem#1\endcsname}%
\let\auto@bib@innerbib\@empty
\bibitem [{\citenamefont {Schr{\"o}dinger}(2025)}]{schrodinger2025life}%
  \BibitemOpen
  \bibfield  {author} {\bibinfo {author} {\bibfnamefont {Erwin}\ \bibnamefont {Schr{\"o}dinger}},\ }\href@noop {} {\emph {\bibinfo {title} {What is life? The physical aspect of the living cell}}}\ (\bibinfo  {publisher} {Rare Treasure Editions},\ \bibinfo {year} {2025})\BibitemShut {NoStop}%
\bibitem [{\citenamefont {te~Vrugt}(2026)}]{teVrugt2026artificial}%
  \BibitemOpen
  \bibinfo {editor} {\bibfnamefont {Michael}\ \bibnamefont {te~Vrugt}},\ ed.,\ \href {\doibase 10.1007/978-3-032-04129-6} {\emph {\bibinfo {title} {Artificial Intelligence and Intelligent Matter: Nanoscience, Soft Matter, Philosophy}}},\ Machine Intelligence for Materials Science\ (\bibinfo  {publisher} {Springer},\ \bibinfo {year} {2026})\BibitemShut {NoStop}%
\bibitem [{\citenamefont {Baulin}\ \emph {et~al.}(2025)\citenamefont {Baulin}, \citenamefont {Giacometti}, \citenamefont {Fedosov}, \citenamefont {Ebbens}, \citenamefont {Varela-Rosales}, \citenamefont {Feliu}, \citenamefont {Chowdhury}, \citenamefont {Hu}, \citenamefont {F{\"u}chslin}, \citenamefont {Dijkstra} \emph {et~al.}}]{baulin2025intelligent}%
  \BibitemOpen
  \bibfield  {author} {\bibinfo {author} {\bibfnamefont {Vladimir~A}\ \bibnamefont {Baulin}}, \bibinfo {author} {\bibfnamefont {Achille}\ \bibnamefont {Giacometti}}, \bibinfo {author} {\bibfnamefont {Dmitry~A}\ \bibnamefont {Fedosov}}, \bibinfo {author} {\bibfnamefont {Stephen}\ \bibnamefont {Ebbens}}, \bibinfo {author} {\bibfnamefont {Nydia~R}\ \bibnamefont {Varela-Rosales}}, \bibinfo {author} {\bibfnamefont {Neus}\ \bibnamefont {Feliu}}, \bibinfo {author} {\bibfnamefont {Mithun}\ \bibnamefont {Chowdhury}}, \bibinfo {author} {\bibfnamefont {Minghan}\ \bibnamefont {Hu}}, \bibinfo {author} {\bibfnamefont {Rudolf}\ \bibnamefont {F{\"u}chslin}}, \bibinfo {author} {\bibfnamefont {Marjolein}\ \bibnamefont {Dijkstra}},  \emph {et~al.},\ }\bibfield  {title} {\enquote {\bibinfo {title} {Intelligent soft matter: towards embodied intelligence},}\ }\href@noop {} {\bibfield  {journal} {\bibinfo  {journal} {Soft Matter}\ }\textbf {\bibinfo {volume} {21}},\ \bibinfo {pages} {4129--4145} (\bibinfo {year}
  {2025})}\BibitemShut {NoStop}%
\bibitem [{\citenamefont {Kaspar}\ \emph {et~al.}(2021)\citenamefont {Kaspar}, \citenamefont {Ravoo}, \citenamefont {van~der Wiel}, \citenamefont {Wegner},\ and\ \citenamefont {Pernice}}]{kaspar2021rise}%
  \BibitemOpen
  \bibfield  {author} {\bibinfo {author} {\bibfnamefont {Corinna}\ \bibnamefont {Kaspar}}, \bibinfo {author} {\bibfnamefont {Bart~Jan}\ \bibnamefont {Ravoo}}, \bibinfo {author} {\bibfnamefont {Wilfred~G}\ \bibnamefont {van~der Wiel}}, \bibinfo {author} {\bibfnamefont {Seraphine~V}\ \bibnamefont {Wegner}}, \ and\ \bibinfo {author} {\bibfnamefont {Wolfram~HP}\ \bibnamefont {Pernice}},\ }\bibfield  {title} {\enquote {\bibinfo {title} {The rise of intelligent matter},}\ }\href@noop {} {\bibfield  {journal} {\bibinfo  {journal} {Nature}\ }\textbf {\bibinfo {volume} {594}},\ \bibinfo {pages} {345--355} (\bibinfo {year} {2021})}\BibitemShut {NoStop}%
\bibitem [{\citenamefont {Ramaswamy}(2010)}]{ramaswamy2010mechanics}%
  \BibitemOpen
  \bibfield  {author} {\bibinfo {author} {\bibfnamefont {Sriram}\ \bibnamefont {Ramaswamy}},\ }\bibfield  {title} {\enquote {\bibinfo {title} {The mechanics and statistics of active matter},}\ }\href@noop {} {\bibfield  {journal} {\bibinfo  {journal} {Annual Review of Condensed Matter Physics}\ }\textbf {\bibinfo {volume} {1}},\ \bibinfo {pages} {323--345} (\bibinfo {year} {2010})}\BibitemShut {NoStop}%
\bibitem [{\citenamefont {Tka{\v{c}}ik}\ and\ \citenamefont {Bialek}(2016)}]{tkavcik2016information}%
  \BibitemOpen
  \bibfield  {author} {\bibinfo {author} {\bibfnamefont {Ga{\v{s}}per}\ \bibnamefont {Tka{\v{c}}ik}}\ and\ \bibinfo {author} {\bibfnamefont {William}\ \bibnamefont {Bialek}},\ }\bibfield  {title} {\enquote {\bibinfo {title} {Information processing in living systems},}\ }\href@noop {} {\bibfield  {journal} {\bibinfo  {journal} {Annual Review of Condensed Matter Physics}\ }\textbf {\bibinfo {volume} {7}},\ \bibinfo {pages} {89--117} (\bibinfo {year} {2016})}\BibitemShut {NoStop}%
\bibitem [{\citenamefont {Toyabe}\ \emph {et~al.}(2011)\citenamefont {Toyabe}, \citenamefont {Watanabe-Nakayama}, \citenamefont {Okamoto}, \citenamefont {Kudo},\ and\ \citenamefont {Muneyuki}}]{toyabe2011thermodynamic}%
  \BibitemOpen
  \bibfield  {author} {\bibinfo {author} {\bibfnamefont {Shoichi}\ \bibnamefont {Toyabe}}, \bibinfo {author} {\bibfnamefont {Takahiro}\ \bibnamefont {Watanabe-Nakayama}}, \bibinfo {author} {\bibfnamefont {Tetsuaki}\ \bibnamefont {Okamoto}}, \bibinfo {author} {\bibfnamefont {Seishi}\ \bibnamefont {Kudo}}, \ and\ \bibinfo {author} {\bibfnamefont {Eiro}\ \bibnamefont {Muneyuki}},\ }\bibfield  {title} {\enquote {\bibinfo {title} {Thermodynamic efficiency and mechanochemical coupling of f1-atpase},}\ }\href@noop {} {\bibfield  {journal} {\bibinfo  {journal} {Proceedings of the National Academy of Sciences}\ }\textbf {\bibinfo {volume} {108}},\ \bibinfo {pages} {17951--17956} (\bibinfo {year} {2011})}\BibitemShut {NoStop}%
\bibitem [{\citenamefont {Ariga}\ \emph {et~al.}(2021)\citenamefont {Ariga}, \citenamefont {Tateishi}, \citenamefont {Tomishige},\ and\ \citenamefont {Mizuno}}]{ariga2021noise}%
  \BibitemOpen
  \bibfield  {author} {\bibinfo {author} {\bibfnamefont {Takayuki}\ \bibnamefont {Ariga}}, \bibinfo {author} {\bibfnamefont {Keito}\ \bibnamefont {Tateishi}}, \bibinfo {author} {\bibfnamefont {Michio}\ \bibnamefont {Tomishige}}, \ and\ \bibinfo {author} {\bibfnamefont {Daisuke}\ \bibnamefont {Mizuno}},\ }\bibfield  {title} {\enquote {\bibinfo {title} {Noise-induced acceleration of single molecule kinesin-1},}\ }\href@noop {} {\bibfield  {journal} {\bibinfo  {journal} {Physical Review Letters}\ }\textbf {\bibinfo {volume} {127}},\ \bibinfo {pages} {178101} (\bibinfo {year} {2021})}\BibitemShut {NoStop}%
\bibitem [{\citenamefont {Viswanathan}\ \emph {et~al.}(2011)\citenamefont {Viswanathan}, \citenamefont {Da~Luz}, \citenamefont {Raposo},\ and\ \citenamefont {Stanley}}]{viswanathan2011physics}%
  \BibitemOpen
  \bibfield  {author} {\bibinfo {author} {\bibfnamefont {Gandhimohan~M}\ \bibnamefont {Viswanathan}}, \bibinfo {author} {\bibfnamefont {Marcos~GE}\ \bibnamefont {Da~Luz}}, \bibinfo {author} {\bibfnamefont {Ernesto~P}\ \bibnamefont {Raposo}}, \ and\ \bibinfo {author} {\bibfnamefont {H~Eugene}\ \bibnamefont {Stanley}},\ }\href@noop {} {\emph {\bibinfo {title} {The physics of foraging: an introduction to random searches and biological encounters}}}\ (\bibinfo  {publisher} {Cambridge University Press},\ \bibinfo {year} {2011})\BibitemShut {NoStop}%
\bibitem [{\citenamefont {Volpe}\ and\ \citenamefont {Volpe}(2017)}]{volpe2017topography}%
  \BibitemOpen
  \bibfield  {author} {\bibinfo {author} {\bibfnamefont {Giorgio}\ \bibnamefont {Volpe}}\ and\ \bibinfo {author} {\bibfnamefont {Giovanni}\ \bibnamefont {Volpe}},\ }\bibfield  {title} {\enquote {\bibinfo {title} {The topography of the environment alters the optimal search strategy for active particles},}\ }\href@noop {} {\bibfield  {journal} {\bibinfo  {journal} {Proceedings of the National Academy of Sciences}\ }\textbf {\bibinfo {volume} {114}},\ \bibinfo {pages} {11350--11355} (\bibinfo {year} {2017})}\BibitemShut {NoStop}%
\bibitem [{\citenamefont {Berg}\ and\ \citenamefont {Purcell}(1977)}]{berg1977physics}%
  \BibitemOpen
  \bibfield  {author} {\bibinfo {author} {\bibfnamefont {Howard~C}\ \bibnamefont {Berg}}\ and\ \bibinfo {author} {\bibfnamefont {Edward~M}\ \bibnamefont {Purcell}},\ }\bibfield  {title} {\enquote {\bibinfo {title} {Physics of chemoreception},}\ }\href@noop {} {\bibfield  {journal} {\bibinfo  {journal} {Biophysical Journal}\ }\textbf {\bibinfo {volume} {20}},\ \bibinfo {pages} {193--219} (\bibinfo {year} {1977})}\BibitemShut {NoStop}%
\bibitem [{\citenamefont {Schnitzer}(1993)}]{schnitzer1993theory}%
  \BibitemOpen
  \bibfield  {author} {\bibinfo {author} {\bibfnamefont {Mark~J}\ \bibnamefont {Schnitzer}},\ }\bibfield  {title} {\enquote {\bibinfo {title} {Theory of continuum random walks and application to chemotaxis},}\ }\href@noop {} {\bibfield  {journal} {\bibinfo  {journal} {Physical Review E}\ }\textbf {\bibinfo {volume} {48}},\ \bibinfo {pages} {2553} (\bibinfo {year} {1993})}\BibitemShut {NoStop}%
\bibitem [{\citenamefont {De~Gennes}(2004)}]{de2004chemotaxis}%
  \BibitemOpen
  \bibfield  {author} {\bibinfo {author} {\bibfnamefont {P-G}\ \bibnamefont {De~Gennes}},\ }\bibfield  {title} {\enquote {\bibinfo {title} {Chemotaxis: the role of internal delays},}\ }\href@noop {} {\bibfield  {journal} {\bibinfo  {journal} {European Biophysics Journal}\ }\textbf {\bibinfo {volume} {33}},\ \bibinfo {pages} {691--693} (\bibinfo {year} {2004})}\BibitemShut {NoStop}%
\bibitem [{\citenamefont {Tsang}\ \emph {et~al.}(2020)\citenamefont {Tsang}, \citenamefont {Demir}, \citenamefont {Ding},\ and\ \citenamefont {Pak}}]{tsang2020roads}%
  \BibitemOpen
  \bibfield  {author} {\bibinfo {author} {\bibfnamefont {Alan~CH}\ \bibnamefont {Tsang}}, \bibinfo {author} {\bibfnamefont {Ebru}\ \bibnamefont {Demir}}, \bibinfo {author} {\bibfnamefont {Yang}\ \bibnamefont {Ding}}, \ and\ \bibinfo {author} {\bibfnamefont {On~Shun}\ \bibnamefont {Pak}},\ }\bibfield  {title} {\enquote {\bibinfo {title} {Roads to smart artificial microswimmers},}\ }\href@noop {} {\bibfield  {journal} {\bibinfo  {journal} {Advanced Intelligent Systems}\ }\textbf {\bibinfo {volume} {2}},\ \bibinfo {pages} {1900137} (\bibinfo {year} {2020})}\BibitemShut {NoStop}%
\bibitem [{\citenamefont {Goh}\ \emph {et~al.}(2022)\citenamefont {Goh}, \citenamefont {Winkler},\ and\ \citenamefont {Gompper}}]{goh2022noisy}%
  \BibitemOpen
  \bibfield  {author} {\bibinfo {author} {\bibfnamefont {Segun}\ \bibnamefont {Goh}}, \bibinfo {author} {\bibfnamefont {Roland~G}\ \bibnamefont {Winkler}}, \ and\ \bibinfo {author} {\bibfnamefont {Gerhard}\ \bibnamefont {Gompper}},\ }\bibfield  {title} {\enquote {\bibinfo {title} {Noisy pursuit and pattern formation of self-steering active particles},}\ }\href@noop {} {\bibfield  {journal} {\bibinfo  {journal} {New Journal of Physics}\ }\textbf {\bibinfo {volume} {24}},\ \bibinfo {pages} {093039} (\bibinfo {year} {2022})}\BibitemShut {NoStop}%
\bibitem [{\citenamefont {Negi}\ \emph {et~al.}(2025)\citenamefont {Negi}, \citenamefont {G~Winkler},\ and\ \citenamefont {Gompper}}]{negi2025binary}%
  \BibitemOpen
  \bibfield  {author} {\bibinfo {author} {\bibfnamefont {Rajendra~Singh}\ \bibnamefont {Negi}}, \bibinfo {author} {\bibfnamefont {Roland}\ \bibnamefont {G~Winkler}}, \ and\ \bibinfo {author} {\bibfnamefont {Gerhard}\ \bibnamefont {Gompper}},\ }\bibfield  {title} {\enquote {\bibinfo {title} {Binary mixtures of intelligent active brownian particles with visual perception},}\ }\href@noop {} {\bibfield  {journal} {\bibinfo  {journal} {New Journal of Physics}\ }\textbf {\bibinfo {volume} {27}},\ \bibinfo {pages} {103301} (\bibinfo {year} {2025})}\BibitemShut {NoStop}%
\bibitem [{\citenamefont {Rode}\ \emph {et~al.}(2024)\citenamefont {Rode}, \citenamefont {Novak},\ and\ \citenamefont {Friedrich}}]{rode2024information}%
  \BibitemOpen
  \bibfield  {author} {\bibinfo {author} {\bibfnamefont {Julian}\ \bibnamefont {Rode}}, \bibinfo {author} {\bibfnamefont {Maja}\ \bibnamefont {Novak}}, \ and\ \bibinfo {author} {\bibfnamefont {Benjamin~M}\ \bibnamefont {Friedrich}},\ }\bibfield  {title} {\enquote {\bibinfo {title} {Information theory of chemotactic agents using both spatial and temporal gradient sensing},}\ }\href@noop {} {\bibfield  {journal} {\bibinfo  {journal} {Physical Review X Life}\ }\textbf {\bibinfo {volume} {2}},\ \bibinfo {pages} {023012} (\bibinfo {year} {2024})}\BibitemShut {NoStop}%
\bibitem [{\citenamefont {Cocconi}\ \emph {et~al.}(2025)\citenamefont {Cocconi}, \citenamefont {Mahault},\ and\ \citenamefont {Piro}}]{cocconi2025dissipation}%
  \BibitemOpen
  \bibfield  {author} {\bibinfo {author} {\bibfnamefont {Luca}\ \bibnamefont {Cocconi}}, \bibinfo {author} {\bibfnamefont {Beno{\^\i}t}\ \bibnamefont {Mahault}}, \ and\ \bibinfo {author} {\bibfnamefont {Lorenzo}\ \bibnamefont {Piro}},\ }\bibfield  {title} {\enquote {\bibinfo {title} {Dissipation-accuracy tradeoffs in autonomous control of smart active matter},}\ }\href@noop {} {\bibfield  {journal} {\bibinfo  {journal} {New Journal of Physics}\ }\textbf {\bibinfo {volume} {27}},\ \bibinfo {pages} {013002} (\bibinfo {year} {2025})}\BibitemShut {NoStop}%
\bibitem [{\citenamefont {Hou}\ \emph {et~al.}(2025)\citenamefont {Hou}, \citenamefont {Zhang}, \citenamefont {Li}, \citenamefont {Yasuda},\ and\ \citenamefont {Komura}}]{hou2025ornstein}%
  \BibitemOpen
  \bibfield  {author} {\bibinfo {author} {\bibfnamefont {Zhanglin}\ \bibnamefont {Hou}}, \bibinfo {author} {\bibfnamefont {Ziluo}\ \bibnamefont {Zhang}}, \bibinfo {author} {\bibfnamefont {Jun}\ \bibnamefont {Li}}, \bibinfo {author} {\bibfnamefont {Kento}\ \bibnamefont {Yasuda}}, \ and\ \bibinfo {author} {\bibfnamefont {Shigeyuki}\ \bibnamefont {Komura}},\ }\bibfield  {title} {\enquote {\bibinfo {title} {Ornstein-Uhlenbeck information swimmers with external and internal feedback controls},}\ }\href@noop {} {\bibfield  {journal} {\bibinfo  {journal} {Europhysics Letters}\ }\textbf {\bibinfo {volume} {150}},\ \bibinfo {pages} {27001} (\bibinfo {year} {2025})}\BibitemShut {NoStop}%
\bibitem [{\citenamefont {Sinha}\ \emph {et~al.}(2025)\citenamefont {Sinha}, \citenamefont {Jangid}, \citenamefont {Sadhu},\ and\ \citenamefont {Ghosh}}]{sinha2025optimal}%
  \BibitemOpen
  \bibfield  {author} {\bibinfo {author} {\bibfnamefont {Abhijit}\ \bibnamefont {Sinha}}, \bibinfo {author} {\bibfnamefont {Sandeep}\ \bibnamefont {Jangid}}, \bibinfo {author} {\bibfnamefont {Tridib}\ \bibnamefont {Sadhu}}, \ and\ \bibinfo {author} {\bibfnamefont {Shankar}\ \bibnamefont {Ghosh}},\ }\bibfield  {title} {\enquote {\bibinfo {title} {Optimal navigation in a noisy environment},}\ }\href@noop {} {\bibfield  {journal} {\bibinfo  {journal} {arXiv preprint arXiv:2512.20336}\ } (\bibinfo {year} {2025})}\BibitemShut {NoStop}%
\bibitem [{\citenamefont {Mori}\ and\ \citenamefont {Mahadevan}(2025)}]{mori2025optimal}%
  \BibitemOpen
  \bibfield  {author} {\bibinfo {author} {\bibfnamefont {Francesco}\ \bibnamefont {Mori}}\ and\ \bibinfo {author} {\bibfnamefont {L}~\bibnamefont {Mahadevan}},\ }\bibfield  {title} {\enquote {\bibinfo {title} {Optimal switching strategies for navigation in stochastic settings},}\ }\href@noop {} {\bibfield  {journal} {\bibinfo  {journal} {Journal of the Royal Society Interface}\ }\textbf {\bibinfo {volume} {22}},\ \bibinfo {pages} {20240677} (\bibinfo {year} {2025})}\BibitemShut {NoStop}%
\bibitem [{\citenamefont {Del~Vecchio}\ \emph {et~al.}(2025)\citenamefont {Del~Vecchio}, \citenamefont {Kulkarni}, \citenamefont {Majumdar},\ and\ \citenamefont {Sabhapandit}}]{del2025proxitaxis}%
  \BibitemOpen
  \bibfield  {author} {\bibinfo {author} {\bibfnamefont {Giuseppe Del~Vecchio}\ \bibnamefont {Del~Vecchio}}, \bibinfo {author} {\bibfnamefont {Manas}\ \bibnamefont {Kulkarni}}, \bibinfo {author} {\bibfnamefont {Satya~N}\ \bibnamefont {Majumdar}}, \ and\ \bibinfo {author} {\bibfnamefont {Sanjib}\ \bibnamefont {Sabhapandit}},\ }\bibfield  {title} {\enquote {\bibinfo {title} {Proxitaxis: an adaptive search strategy based on proximity and stochastic resetting},}\ }\href@noop {} {\bibfield  {journal} {\bibinfo  {journal} {arXiv preprint arXiv:2507.05800}\ } (\bibinfo {year} {2025})}\BibitemShut {NoStop}%
\bibitem [{\citenamefont {Song}\ \emph {et~al.}(2026)\citenamefont {Song}, \citenamefont {Shao}, \citenamefont {Zhu}, \citenamefont {Yang}, \citenamefont {He}, \citenamefont {Komura},\ and\ \citenamefont {Hou}}]{song2026ornsteinuhlenbeckinformationparticlenew}%
  \BibitemOpen
  \bibfield  {author} {\bibinfo {author} {\bibfnamefont {Xin}\ \bibnamefont {Song}}, \bibinfo {author} {\bibfnamefont {Xiji}\ \bibnamefont {Shao}}, \bibinfo {author} {\bibfnamefont {Yanwen}\ \bibnamefont {Zhu}}, \bibinfo {author} {\bibfnamefont {Cheng}\ \bibnamefont {Yang}}, \bibinfo {author} {\bibfnamefont {Linli}\ \bibnamefont {He}}, \bibinfo {author} {\bibfnamefont {Shigeyuki}\ \bibnamefont {Komura}}, \ and\ \bibinfo {author} {\bibfnamefont {Zhanglin}\ \bibnamefont {Hou}},\ }\href {https://arxiv.org/abs/2602.06340} {\enquote {\bibinfo {title} {Ornstein-Uhlenbeck information particle: A new candidate of active agent},}\ } (\bibinfo {year} {2026}),\ \Eprint {http://arxiv.org/abs/2602.06340} {arXiv:2602.06340 [cond-mat.stat-mech]} \BibitemShut {NoStop}%
\bibitem [{\citenamefont {Colabrese}\ \emph {et~al.}(2017)\citenamefont {Colabrese}, \citenamefont {Gustavsson}, \citenamefont {Celani},\ and\ \citenamefont {Biferale}}]{colabrese2017flow}%
  \BibitemOpen
  \bibfield  {author} {\bibinfo {author} {\bibfnamefont {Simona}\ \bibnamefont {Colabrese}}, \bibinfo {author} {\bibfnamefont {Kristian}\ \bibnamefont {Gustavsson}}, \bibinfo {author} {\bibfnamefont {Antonio}\ \bibnamefont {Celani}}, \ and\ \bibinfo {author} {\bibfnamefont {Luca}\ \bibnamefont {Biferale}},\ }\bibfield  {title} {\enquote {\bibinfo {title} {Flow navigation by smart microswimmers via reinforcement learning},}\ }\href@noop {} {\bibfield  {journal} {\bibinfo  {journal} {Physical Review Letters}\ }\textbf {\bibinfo {volume} {118}},\ \bibinfo {pages} {158004} (\bibinfo {year} {2017})}\BibitemShut {NoStop}%
\bibitem [{\citenamefont {Jacob}\ \emph {et~al.}(2025)\citenamefont {Jacob}, \citenamefont {Mohapatra}, \citenamefont {A}, \citenamefont {Mathew},\ and\ \citenamefont {Sinha~Mahapatra}}]{jacob2025mixing}%
  \BibitemOpen
  \bibfield  {author} {\bibinfo {author} {\bibfnamefont {Thomas}\ \bibnamefont {Jacob}}, \bibinfo {author} {\bibfnamefont {Siddhant}\ \bibnamefont {Mohapatra}}, \bibinfo {author} {\bibfnamefont {Rajalingam}\ \bibnamefont {A}}, \bibinfo {author} {\bibfnamefont {Sam}\ \bibnamefont {Mathew}}, \ and\ \bibinfo {author} {\bibfnamefont {Pallab}\ \bibnamefont {Sinha~Mahapatra}},\ }\bibfield  {title} {\enquote {\bibinfo {title} {Mixing of a binary passive particle system using smart active particles},}\ }\href@noop {} {\bibfield  {journal} {\bibinfo  {journal} {Scientific Reports}\ } (\bibinfo {year} {2025})}\BibitemShut {NoStop}%
\bibitem [{\citenamefont {Gassner}\ \emph {et~al.}(2023)\citenamefont {Gassner}, \citenamefont {Goh}, \citenamefont {Gompper},\ and\ \citenamefont {Winkler}}]{gassner2023noisy}%
  \BibitemOpen
  \bibfield  {author} {\bibinfo {author} {\bibfnamefont {Marielle}\ \bibnamefont {Gassner}}, \bibinfo {author} {\bibfnamefont {Segun}\ \bibnamefont {Goh}}, \bibinfo {author} {\bibfnamefont {Gerhard}\ \bibnamefont {Gompper}}, \ and\ \bibinfo {author} {\bibfnamefont {Roland~G}\ \bibnamefont {Winkler}},\ }\bibfield  {title} {\enquote {\bibinfo {title} {Noisy pursuit by a self-steering active particle in confinement},}\ }\href@noop {} {\bibfield  {journal} {\bibinfo  {journal} {Europhysics Letters}\ }\textbf {\bibinfo {volume} {142}},\ \bibinfo {pages} {21002} (\bibinfo {year} {2023})}\BibitemShut {NoStop}%
\bibitem [{\citenamefont {Putzke}\ and\ \citenamefont {Stark}(2023)}]{putzke2023optimal}%
  \BibitemOpen
  \bibfield  {author} {\bibinfo {author} {\bibfnamefont {Mischa}\ \bibnamefont {Putzke}}\ and\ \bibinfo {author} {\bibfnamefont {Holger}\ \bibnamefont {Stark}},\ }\bibfield  {title} {\enquote {\bibinfo {title} {Optimal navigation of a smart active particle: directional and distance sensing},}\ }\href@noop {} {\bibfield  {journal} {\bibinfo  {journal} {The European Physical Journal E}\ }\textbf {\bibinfo {volume} {46}},\ \bibinfo {pages} {48} (\bibinfo {year} {2023})}\BibitemShut {NoStop}%
\bibitem [{\citenamefont {Cichos}\ \emph {et~al.}(2020)\citenamefont {Cichos}, \citenamefont {Gustavsson}, \citenamefont {Mehlig},\ and\ \citenamefont {Volpe}}]{cichos2020machine}%
  \BibitemOpen
  \bibfield  {author} {\bibinfo {author} {\bibfnamefont {Frank}\ \bibnamefont {Cichos}}, \bibinfo {author} {\bibfnamefont {Kristian}\ \bibnamefont {Gustavsson}}, \bibinfo {author} {\bibfnamefont {Bernhard}\ \bibnamefont {Mehlig}}, \ and\ \bibinfo {author} {\bibfnamefont {Giovanni}\ \bibnamefont {Volpe}},\ }\bibfield  {title} {\enquote {\bibinfo {title} {Machine learning for active matter},}\ }\href@noop {} {\bibfield  {journal} {\bibinfo  {journal} {Nature Machine Intelligence}\ }\textbf {\bibinfo {volume} {2}},\ \bibinfo {pages} {94--103} (\bibinfo {year} {2020})}\BibitemShut {NoStop}%
\bibitem [{\citenamefont {Gustavsson}\ \emph {et~al.}(2017)\citenamefont {Gustavsson}, \citenamefont {Biferale}, \citenamefont {Celani},\ and\ \citenamefont {Colabrese}}]{gustavsson2017finding}%
  \BibitemOpen
  \bibfield  {author} {\bibinfo {author} {\bibfnamefont {Kristian}\ \bibnamefont {Gustavsson}}, \bibinfo {author} {\bibfnamefont {Luca}\ \bibnamefont {Biferale}}, \bibinfo {author} {\bibfnamefont {Antonio}\ \bibnamefont {Celani}}, \ and\ \bibinfo {author} {\bibfnamefont {Simona}\ \bibnamefont {Colabrese}},\ }\bibfield  {title} {\enquote {\bibinfo {title} {Finding efficient swimming strategies in a three-dimensional chaotic flow by reinforcement learning},}\ }\href@noop {} {\bibfield  {journal} {\bibinfo  {journal} {The European Physical Journal E}\ }\textbf {\bibinfo {volume} {40}},\ \bibinfo {pages} {1--6} (\bibinfo {year} {2017})}\BibitemShut {NoStop}%
\bibitem [{\citenamefont {Schneider}\ and\ \citenamefont {Stark}(2019)}]{schneider2019optimal}%
  \BibitemOpen
  \bibfield  {author} {\bibinfo {author} {\bibfnamefont {Elias}\ \bibnamefont {Schneider}}\ and\ \bibinfo {author} {\bibfnamefont {Holger}\ \bibnamefont {Stark}},\ }\bibfield  {title} {\enquote {\bibinfo {title} {Optimal steering of a smart active particle},}\ }\href@noop {} {\bibfield  {journal} {\bibinfo  {journal} {Europhysics Letters}\ }\textbf {\bibinfo {volume} {127}},\ \bibinfo {pages} {64003} (\bibinfo {year} {2019})}\BibitemShut {NoStop}%
\bibitem [{\citenamefont {Durve}\ \emph {et~al.}(2020)\citenamefont {Durve}, \citenamefont {Peruani},\ and\ \citenamefont {Celani}}]{durve2020learning}%
  \BibitemOpen
  \bibfield  {author} {\bibinfo {author} {\bibfnamefont {Mihir}\ \bibnamefont {Durve}}, \bibinfo {author} {\bibfnamefont {Fernando}\ \bibnamefont {Peruani}}, \ and\ \bibinfo {author} {\bibfnamefont {Antonio}\ \bibnamefont {Celani}},\ }\bibfield  {title} {\enquote {\bibinfo {title} {Learning to flock through reinforcement},}\ }\href@noop {} {\bibfield  {journal} {\bibinfo  {journal} {Physical Review E}\ }\textbf {\bibinfo {volume} {102}},\ \bibinfo {pages} {012601} (\bibinfo {year} {2020})}\BibitemShut {NoStop}%
\bibitem [{\citenamefont {Nasiri}\ \emph {et~al.}(2023)\citenamefont {Nasiri}, \citenamefont {L{\"o}wen},\ and\ \citenamefont {Liebchen}}]{nasiri2023optimal}%
  \BibitemOpen
  \bibfield  {author} {\bibinfo {author} {\bibfnamefont {Mahdi}\ \bibnamefont {Nasiri}}, \bibinfo {author} {\bibfnamefont {Hartmut}\ \bibnamefont {L{\"o}wen}}, \ and\ \bibinfo {author} {\bibfnamefont {Benno}\ \bibnamefont {Liebchen}},\ }\bibfield  {title} {\enquote {\bibinfo {title} {Optimal active particle navigation meets machine learning (a)},}\ }\href@noop {} {\bibfield  {journal} {\bibinfo  {journal} {Europhysics Letters}\ }\textbf {\bibinfo {volume} {142}},\ \bibinfo {pages} {17001} (\bibinfo {year} {2023})}\BibitemShut {NoStop}%
\bibitem [{\citenamefont {Nasiri}\ \emph {et~al.}(2024)\citenamefont {Nasiri}, \citenamefont {Loran},\ and\ \citenamefont {Liebchen}}]{nasiri2024smart}%
  \BibitemOpen
  \bibfield  {author} {\bibinfo {author} {\bibfnamefont {Mahdi}\ \bibnamefont {Nasiri}}, \bibinfo {author} {\bibfnamefont {Edwin}\ \bibnamefont {Loran}}, \ and\ \bibinfo {author} {\bibfnamefont {Benno}\ \bibnamefont {Liebchen}},\ }\bibfield  {title} {\enquote {\bibinfo {title} {Smart active particles learn and transcend bacterial foraging strategies},}\ }\href@noop {} {\bibfield  {journal} {\bibinfo  {journal} {Proceedings of the National Academy of Sciences}\ }\textbf {\bibinfo {volume} {121}},\ \bibinfo {pages} {e2317618121} (\bibinfo {year} {2024})}\BibitemShut {NoStop}%
\bibitem [{\citenamefont {Heinonen}\ \emph {et~al.}(2025)\citenamefont {Heinonen}, \citenamefont {Biferale}, \citenamefont {Celani},\ and\ \citenamefont {Vergassola}}]{heinonen2025optimal}%
  \BibitemOpen
  \bibfield  {author} {\bibinfo {author} {\bibfnamefont {Robin~A}\ \bibnamefont {Heinonen}}, \bibinfo {author} {\bibfnamefont {Luca}\ \bibnamefont {Biferale}}, \bibinfo {author} {\bibfnamefont {Antonio}\ \bibnamefont {Celani}}, \ and\ \bibinfo {author} {\bibfnamefont {Massimo}\ \bibnamefont {Vergassola}},\ }\bibfield  {title} {\enquote {\bibinfo {title} {Optimal trajectories for bayesian olfactory search in turbulent flows: The low information limit and beyond},}\ }\href@noop {} {\bibfield  {journal} {\bibinfo  {journal} {Physical Review Fluids}\ }\textbf {\bibinfo {volume} {10}},\ \bibinfo {pages} {044601} (\bibinfo {year} {2025})}\BibitemShut {NoStop}%
\bibitem [{\citenamefont {Singh}\ \emph {et~al.}(2026)\citenamefont {Singh}, \citenamefont {Jena}, \citenamefont {Kumar},\ and\ \citenamefont {Mishra}}]{singh2026homingreinforcementlearning}%
  \BibitemOpen
  \bibfield  {author} {\bibinfo {author} {\bibfnamefont {Riya}\ \bibnamefont {Singh}}, \bibinfo {author} {\bibfnamefont {Pratikshya}\ \bibnamefont {Jena}}, \bibinfo {author} {\bibfnamefont {Anish}\ \bibnamefont {Kumar}}, \ and\ \bibinfo {author} {\bibfnamefont {Shradha}\ \bibnamefont {Mishra}},\ }\href {https://arxiv.org/abs/2602.08566} {\enquote {\bibinfo {title} {Homing through reinforcement learning},}\ } (\bibinfo {year} {2026}),\ \Eprint {http://arxiv.org/abs/2602.08566} {arXiv:2602.08566 [cond-mat.soft]} \BibitemShut {NoStop}%
\bibitem [{\citenamefont {M{\"u}ller}\ and\ \citenamefont {Wehner}(1988)}]{muller1988path}%
  \BibitemOpen
  \bibfield  {author} {\bibinfo {author} {\bibfnamefont {Martin}\ \bibnamefont {M{\"u}ller}}\ and\ \bibinfo {author} {\bibfnamefont {R{\"u}diger}\ \bibnamefont {Wehner}},\ }\bibfield  {title} {\enquote {\bibinfo {title} {Path integration in desert ants, cataglyphis fortis},}\ }\href@noop {} {\bibfield  {journal} {\bibinfo  {journal} {Proceedings of the National Academy of Sciences}\ }\textbf {\bibinfo {volume} {85}},\ \bibinfo {pages} {5287--5290} (\bibinfo {year} {1988})}\BibitemShut {NoStop}%
\bibitem [{\citenamefont {Hartmann}\ and\ \citenamefont {Wehner}(1995)}]{hartmann1995ant}%
  \BibitemOpen
  \bibfield  {author} {\bibinfo {author} {\bibfnamefont {Georg}\ \bibnamefont {Hartmann}}\ and\ \bibinfo {author} {\bibfnamefont {R{\"u}diger}\ \bibnamefont {Wehner}},\ }\bibfield  {title} {\enquote {\bibinfo {title} {The ant's path integration system: a neural architecture},}\ }\href@noop {} {\bibfield  {journal} {\bibinfo  {journal} {Biological Cybernetics}\ }\textbf {\bibinfo {volume} {73}},\ \bibinfo {pages} {483--497} (\bibinfo {year} {1995})}\BibitemShut {NoStop}%
\bibitem [{\citenamefont {Lenz}\ \emph {et~al.}(2012)\citenamefont {Lenz}, \citenamefont {Ings}, \citenamefont {Chittka}, \citenamefont {Chechkin},\ and\ \citenamefont {Klages}}]{lenz2012spatiotemporal}%
  \BibitemOpen
  \bibfield  {author} {\bibinfo {author} {\bibfnamefont {Friedrich}\ \bibnamefont {Lenz}}, \bibinfo {author} {\bibfnamefont {Thomas~C}\ \bibnamefont {Ings}}, \bibinfo {author} {\bibfnamefont {Lars}\ \bibnamefont {Chittka}}, \bibinfo {author} {\bibfnamefont {Aleksei~V}\ \bibnamefont {Chechkin}}, \ and\ \bibinfo {author} {\bibfnamefont {Rainer}\ \bibnamefont {Klages}},\ }\bibfield  {title} {\enquote {\bibinfo {title} {Spatiotemporal dynamics of bumblebees foraging under predation risk},}\ }\href@noop {} {\bibfield  {journal} {\bibinfo  {journal} {Physical Review Letters}\ }\textbf {\bibinfo {volume} {108}},\ \bibinfo {pages} {098103} (\bibinfo {year} {2012})}\BibitemShut {NoStop}%
\bibitem [{\citenamefont {Wiltschko}\ and\ \citenamefont {Wiltschko}(2023)}]{wiltschko2023animal}%
  \BibitemOpen
  \bibfield  {author} {\bibinfo {author} {\bibfnamefont {Roswitha}\ \bibnamefont {Wiltschko}}\ and\ \bibinfo {author} {\bibfnamefont {Wolfgang}\ \bibnamefont {Wiltschko}},\ }\bibfield  {title} {\enquote {\bibinfo {title} {Animal navigation: how animals use environmental factors to find their way},}\ }\href@noop {} {\bibfield  {journal} {\bibinfo  {journal} {The European Physical Journal Special Topics}\ }\textbf {\bibinfo {volume} {232}},\ \bibinfo {pages} {237--252} (\bibinfo {year} {2023})}\BibitemShut {NoStop}%
\bibitem [{\citenamefont {Garrett}(1922)}]{garrett1922study}%
  \BibitemOpen
  \bibfield  {author} {\bibinfo {author} {\bibfnamefont {Henry~Edward}\ \bibnamefont {Garrett}},\ }\href@noop {} {\emph {\bibinfo {title} {A study of the relation of accuracy to speed}}},\ Vol.~\bibinfo {volume} {8}\ (\bibinfo  {publisher} {Columbia University},\ \bibinfo {year} {1922})\BibitemShut {NoStop}%
\bibitem [{\citenamefont {Ratcliff}\ and\ \citenamefont {McKoon}(2008)}]{ratcliff2008diffusion}%
  \BibitemOpen
  \bibfield  {author} {\bibinfo {author} {\bibfnamefont {Roger}\ \bibnamefont {Ratcliff}}\ and\ \bibinfo {author} {\bibfnamefont {Gail}\ \bibnamefont {McKoon}},\ }\bibfield  {title} {\enquote {\bibinfo {title} {The diffusion decision model: theory and data for two-choice decision tasks},}\ }\href@noop {} {\bibfield  {journal} {\bibinfo  {journal} {Neural Computation}\ }\textbf {\bibinfo {volume} {20}},\ \bibinfo {pages} {873--922} (\bibinfo {year} {2008})}\BibitemShut {NoStop}%
\bibitem [{\citenamefont {Ratcliff}\ \emph {et~al.}(2016)\citenamefont {Ratcliff}, \citenamefont {Smith}, \citenamefont {Brown},\ and\ \citenamefont {McKoon}}]{ratcliff2016diffusion}%
  \BibitemOpen
  \bibfield  {author} {\bibinfo {author} {\bibfnamefont {Roger}\ \bibnamefont {Ratcliff}}, \bibinfo {author} {\bibfnamefont {Philip~L}\ \bibnamefont {Smith}}, \bibinfo {author} {\bibfnamefont {Scott~D}\ \bibnamefont {Brown}}, \ and\ \bibinfo {author} {\bibfnamefont {Gail}\ \bibnamefont {McKoon}},\ }\bibfield  {title} {\enquote {\bibinfo {title} {Diffusion decision model: Current issues and history},}\ }\href@noop {} {\bibfield  {journal} {\bibinfo  {journal} {Trends in Cognitive Sciences}\ }\textbf {\bibinfo {volume} {20}},\ \bibinfo {pages} {260--281} (\bibinfo {year} {2016})}\BibitemShut {NoStop}%
\bibitem [{\citenamefont {Siggia}\ and\ \citenamefont {Vergassola}(2013)}]{siggia2013decisions}%
  \BibitemOpen
  \bibfield  {author} {\bibinfo {author} {\bibfnamefont {Eric~D}\ \bibnamefont {Siggia}}\ and\ \bibinfo {author} {\bibfnamefont {Massimo}\ \bibnamefont {Vergassola}},\ }\bibfield  {title} {\enquote {\bibinfo {title} {Decisions on the fly in cellular sensory systems},}\ }\href@noop {} {\bibfield  {journal} {\bibinfo  {journal} {Proceedings of the National Academy of Sciences}\ }\textbf {\bibinfo {volume} {110}},\ \bibinfo {pages} {E3704--E3712} (\bibinfo {year} {2013})}\BibitemShut {NoStop}%
\bibitem [{\citenamefont {Chittka}\ \emph {et~al.}(2003)\citenamefont {Chittka}, \citenamefont {Dyer}, \citenamefont {Bock},\ and\ \citenamefont {Dornhaus}}]{chittka2003bees}%
  \BibitemOpen
  \bibfield  {author} {\bibinfo {author} {\bibfnamefont {Lars}\ \bibnamefont {Chittka}}, \bibinfo {author} {\bibfnamefont {Adrian~G}\ \bibnamefont {Dyer}}, \bibinfo {author} {\bibfnamefont {Fiola}\ \bibnamefont {Bock}}, \ and\ \bibinfo {author} {\bibfnamefont {Anna}\ \bibnamefont {Dornhaus}},\ }\bibfield  {title} {\enquote {\bibinfo {title} {Bees trade off foraging speed for accuracy},}\ }\href@noop {} {\bibfield  {journal} {\bibinfo  {journal} {Nature}\ }\textbf {\bibinfo {volume} {424}},\ \bibinfo {pages} {388--388} (\bibinfo {year} {2003})}\BibitemShut {NoStop}%
\bibitem [{\citenamefont {Durmaz}\ \emph {et~al.}(2023)\citenamefont {Durmaz}, \citenamefont {Sarmiento}, \citenamefont {Fortunato}, \citenamefont {Das}, \citenamefont {Diamond}, \citenamefont {Bueti},\ and\ \citenamefont {Rold{\'a}n}}]{durmaz2023human}%
  \BibitemOpen
  \bibfield  {author} {\bibinfo {author} {\bibfnamefont {Ayb{\"u}ke}\ \bibnamefont {Durmaz}}, \bibinfo {author} {\bibfnamefont {Yonathan}\ \bibnamefont {Sarmiento}}, \bibinfo {author} {\bibfnamefont {Gianfranco}\ \bibnamefont {Fortunato}}, \bibinfo {author} {\bibfnamefont {Debraj}\ \bibnamefont {Das}}, \bibinfo {author} {\bibfnamefont {Mathew~Ernst}\ \bibnamefont {Diamond}}, \bibinfo {author} {\bibfnamefont {Domenica}\ \bibnamefont {Bueti}}, \ and\ \bibinfo {author} {\bibfnamefont {{\'E}dgar}\ \bibnamefont {Rold{\'a}n}},\ }\bibfield  {title} {\enquote {\bibinfo {title} {Human perceptual decision making of nonequilibrium fluctuations},}\ }\href@noop {} {\bibfield  {journal} {\bibinfo  {journal} {arXiv preprint arXiv:2311.12692}\ } (\bibinfo {year} {2023})}\BibitemShut {NoStop}%
\bibitem [{\citenamefont {Brunton}\ \emph {et~al.}(2013)\citenamefont {Brunton}, \citenamefont {Botvinick},\ and\ \citenamefont {Brody}}]{brunton2013rats}%
  \BibitemOpen
  \bibfield  {author} {\bibinfo {author} {\bibfnamefont {Bingni~W}\ \bibnamefont {Brunton}}, \bibinfo {author} {\bibfnamefont {Matthew~M}\ \bibnamefont {Botvinick}}, \ and\ \bibinfo {author} {\bibfnamefont {Carlos~D}\ \bibnamefont {Brody}},\ }\bibfield  {title} {\enquote {\bibinfo {title} {Rats and humans can optimally accumulate evidence for decision-making},}\ }\href@noop {} {\bibfield  {journal} {\bibinfo  {journal} {Science}\ }\textbf {\bibinfo {volume} {340}},\ \bibinfo {pages} {95--98} (\bibinfo {year} {2013})}\BibitemShut {NoStop}%
\bibitem [{\citenamefont {Vergassola}\ \emph {et~al.}(2007)\citenamefont {Vergassola}, \citenamefont {Villermaux},\ and\ \citenamefont {Shraiman}}]{vergassola2007infotaxis}%
  \BibitemOpen
  \bibfield  {author} {\bibinfo {author} {\bibfnamefont {Massimo}\ \bibnamefont {Vergassola}}, \bibinfo {author} {\bibfnamefont {Emmanuel}\ \bibnamefont {Villermaux}}, \ and\ \bibinfo {author} {\bibfnamefont {Boris~I}\ \bibnamefont {Shraiman}},\ }\bibfield  {title} {\enquote {\bibinfo {title} {‘infotaxis’ as a strategy for searching without gradients},}\ }\href@noop {} {\bibfield  {journal} {\bibinfo  {journal} {Nature}\ }\textbf {\bibinfo {volume} {445}},\ \bibinfo {pages} {406--409} (\bibinfo {year} {2007})}\BibitemShut {NoStop}%
\bibitem [{\citenamefont {Celani}\ \emph {et~al.}(2014)\citenamefont {Celani}, \citenamefont {Villermaux},\ and\ \citenamefont {Vergassola}}]{celani2014odor}%
  \BibitemOpen
  \bibfield  {author} {\bibinfo {author} {\bibfnamefont {Antonio}\ \bibnamefont {Celani}}, \bibinfo {author} {\bibfnamefont {Emmanuel}\ \bibnamefont {Villermaux}}, \ and\ \bibinfo {author} {\bibfnamefont {Massimo}\ \bibnamefont {Vergassola}},\ }\bibfield  {title} {\enquote {\bibinfo {title} {Odor landscapes in turbulent environments},}\ }\href@noop {} {\bibfield  {journal} {\bibinfo  {journal} {Physical Review X}\ }\textbf {\bibinfo {volume} {4}},\ \bibinfo {pages} {041015} (\bibinfo {year} {2014})}\BibitemShut {NoStop}%
\bibitem [{\citenamefont {Cover}(1999)}]{cover1999elements}%
  \BibitemOpen
  \bibfield  {author} {\bibinfo {author} {\bibfnamefont {Thomas~M}\ \bibnamefont {Cover}},\ }\href@noop {} {\emph {\bibinfo {title} {Elements of information theory}}}\ (\bibinfo  {publisher} {John Wiley \& Sons},\ \bibinfo {year} {1999})\BibitemShut {NoStop}%
\bibitem [{\citenamefont {Kullback}(1997)}]{kullback1997information}%
  \BibitemOpen
  \bibfield  {author} {\bibinfo {author} {\bibfnamefont {Solomon}\ \bibnamefont {Kullback}},\ }\href@noop {} {\emph {\bibinfo {title} {Information theory and statistics}}}\ (\bibinfo  {publisher} {Courier Corporation},\ \bibinfo {year} {1997})\BibitemShut {NoStop}%
\bibitem [{\citenamefont {Howse}\ \emph {et~al.}(2007)\citenamefont {Howse}, \citenamefont {Jones}, \citenamefont {Ryan}, \citenamefont {Gough}, \citenamefont {Vafabakhsh},\ and\ \citenamefont {Golestanian}}]{howse2007self}%
  \BibitemOpen
  \bibfield  {author} {\bibinfo {author} {\bibfnamefont {Jonathan~R}\ \bibnamefont {Howse}}, \bibinfo {author} {\bibfnamefont {Richard~AL}\ \bibnamefont {Jones}}, \bibinfo {author} {\bibfnamefont {Anthony~J}\ \bibnamefont {Ryan}}, \bibinfo {author} {\bibfnamefont {Tim}\ \bibnamefont {Gough}}, \bibinfo {author} {\bibfnamefont {Reza}\ \bibnamefont {Vafabakhsh}}, \ and\ \bibinfo {author} {\bibfnamefont {Ramin}\ \bibnamefont {Golestanian}},\ }\bibfield  {title} {\enquote {\bibinfo {title} {Self-motile colloidal particles: from directed propulsion to random walk},}\ }\href@noop {} {\bibfield  {journal} {\bibinfo  {journal} {Physical Review Letters}\ }\textbf {\bibinfo {volume} {99}},\ \bibinfo {pages} {048102} (\bibinfo {year} {2007})}\BibitemShut {NoStop}%
\bibitem [{\citenamefont {Te~Vrugt}\ and\ \citenamefont {Wittkowski}(2025)}]{te2025metareview}%
  \BibitemOpen
  \bibfield  {author} {\bibinfo {author} {\bibfnamefont {Michael}\ \bibnamefont {Te~Vrugt}}\ and\ \bibinfo {author} {\bibfnamefont {Raphael}\ \bibnamefont {Wittkowski}},\ }\bibfield  {title} {\enquote {\bibinfo {title} {Metareview: a survey of active matter reviews},}\ }\href@noop {} {\bibfield  {journal} {\bibinfo  {journal} {The European Physical Journal E}\ }\textbf {\bibinfo {volume} {48}},\ \bibinfo {pages} {12} (\bibinfo {year} {2025})}\BibitemShut {NoStop}%
\bibitem [{\citenamefont {Card{\'e}}\ and\ \citenamefont {Willis}(2008)}]{carde2008navigational}%
  \BibitemOpen
  \bibfield  {author} {\bibinfo {author} {\bibfnamefont {Ring~T}\ \bibnamefont {Card{\'e}}}\ and\ \bibinfo {author} {\bibfnamefont {Mark~A}\ \bibnamefont {Willis}},\ }\bibfield  {title} {\enquote {\bibinfo {title} {Navigational strategies used by insects to find distant, wind-borne sources of odor},}\ }\href@noop {} {\bibfield  {journal} {\bibinfo  {journal} {Journal of Chemical Ecology}\ }\textbf {\bibinfo {volume} {34}},\ \bibinfo {pages} {854--866} (\bibinfo {year} {2008})}\BibitemShut {NoStop}%
\bibitem [{\citenamefont {Byrne}\ \emph {et~al.}(2003)\citenamefont {Byrne}, \citenamefont {Dacke}, \citenamefont {Nordstr{\"o}m}, \citenamefont {Scholtz},\ and\ \citenamefont {Warrant}}]{byrne2003visual}%
  \BibitemOpen
  \bibfield  {author} {\bibinfo {author} {\bibfnamefont {Marcus}\ \bibnamefont {Byrne}}, \bibinfo {author} {\bibfnamefont {Marie}\ \bibnamefont {Dacke}}, \bibinfo {author} {\bibfnamefont {Peter}\ \bibnamefont {Nordstr{\"o}m}}, \bibinfo {author} {\bibfnamefont {Clarke}\ \bibnamefont {Scholtz}}, \ and\ \bibinfo {author} {\bibfnamefont {Eric}\ \bibnamefont {Warrant}},\ }\bibfield  {title} {\enquote {\bibinfo {title} {Visual cues used by ball-rolling dung beetles for orientation},}\ }\href@noop {} {\bibfield  {journal} {\bibinfo  {journal} {Journal of Comparative Physiology A}\ }\textbf {\bibinfo {volume} {189}},\ \bibinfo {pages} {411--418} (\bibinfo {year} {2003})}\BibitemShut {NoStop}%
\bibitem [{\citenamefont {Dacke}\ \emph {et~al.}(2013)\citenamefont {Dacke}, \citenamefont {Baird}, \citenamefont {Byrne}, \citenamefont {Scholtz},\ and\ \citenamefont {Warrant}}]{dacke2013dung}%
  \BibitemOpen
  \bibfield  {author} {\bibinfo {author} {\bibfnamefont {Marie}\ \bibnamefont {Dacke}}, \bibinfo {author} {\bibfnamefont {Emily}\ \bibnamefont {Baird}}, \bibinfo {author} {\bibfnamefont {Marcus}\ \bibnamefont {Byrne}}, \bibinfo {author} {\bibfnamefont {Clarke~H}\ \bibnamefont {Scholtz}}, \ and\ \bibinfo {author} {\bibfnamefont {Eric~J}\ \bibnamefont {Warrant}},\ }\bibfield  {title} {\enquote {\bibinfo {title} {Dung beetles use the milky way for orientation},}\ }\href@noop {} {\bibfield  {journal} {\bibinfo  {journal} {Current Biology}\ }\textbf {\bibinfo {volume} {23}},\ \bibinfo {pages} {298--300} (\bibinfo {year} {2013})}\BibitemShut {NoStop}%
\bibitem [{\citenamefont {Kumar}\ \emph {et~al.}(2020)\citenamefont {Kumar}, \citenamefont {Sadekar},\ and\ \citenamefont {Basu}}]{kumar2020active}%
  \BibitemOpen
  \bibfield  {author} {\bibinfo {author} {\bibfnamefont {Vijay}\ \bibnamefont {Kumar}}, \bibinfo {author} {\bibfnamefont {Onkar}\ \bibnamefont {Sadekar}}, \ and\ \bibinfo {author} {\bibfnamefont {Urna}\ \bibnamefont {Basu}},\ }\bibfield  {title} {\enquote {\bibinfo {title} {Active Brownian motion in two dimensions under stochastic resetting},}\ }\href@noop {} {\bibfield  {journal} {\bibinfo  {journal} {Physical Review E}\ }\textbf {\bibinfo {volume} {102}},\ \bibinfo {pages} {052129} (\bibinfo {year} {2020})}\BibitemShut {NoStop}%
\bibitem [{\citenamefont {Baouche}\ \emph {et~al.}(2024)\citenamefont {Baouche}, \citenamefont {Franosch}, \citenamefont {Meiners},\ and\ \citenamefont {Kurzthaler}}]{baouche2024active}%
  \BibitemOpen
  \bibfield  {author} {\bibinfo {author} {\bibfnamefont {Yanis}\ \bibnamefont {Baouche}}, \bibinfo {author} {\bibfnamefont {Thomas}\ \bibnamefont {Franosch}}, \bibinfo {author} {\bibfnamefont {Matthias}\ \bibnamefont {Meiners}}, \ and\ \bibinfo {author} {\bibfnamefont {Christina}\ \bibnamefont {Kurzthaler}},\ }\bibfield  {title} {\enquote {\bibinfo {title} {Active Brownian particle under stochastic orientational resetting},}\ }\href@noop {} {\bibfield  {journal} {\bibinfo  {journal} {New Journal of Physics}\ }\textbf {\bibinfo {volume} {26}},\ \bibinfo {pages} {073041} (\bibinfo {year} {2024})}\BibitemShut {NoStop}%
\bibitem [{\citenamefont {Olsen}\ and\ \citenamefont {L{\"o}wen}(2024)}]{olsen2024optimal}%
  \BibitemOpen
  \bibfield  {author} {\bibinfo {author} {\bibfnamefont {Kristian~St{\o}levik}\ \bibnamefont {Olsen}}\ and\ \bibinfo {author} {\bibfnamefont {Hartmut}\ \bibnamefont {L{\"o}wen}},\ }\bibfield  {title} {\enquote {\bibinfo {title} {Optimal diffusion of chiral active particles with strategic reorientations},}\ }\href@noop {} {\bibfield  {journal} {\bibinfo  {journal} {Physical Review E}\ }\textbf {\bibinfo {volume} {110}},\ \bibinfo {pages} {064606} (\bibinfo {year} {2024})}\BibitemShut {NoStop}%
\bibitem [{\citenamefont {Shee}(2025)}]{shee2025steeringchiralactivebrownian}%
  \BibitemOpen
  \bibfield  {author} {\bibinfo {author} {\bibfnamefont {Amir}\ \bibnamefont {Shee}},\ }\href {https://arxiv.org/abs/2508.12223} {\enquote {\bibinfo {title} {Steering chiral active brownian motion via stochastic position-orientation resetting},}\ } (\bibinfo {year} {2025}),\ \Eprint {http://arxiv.org/abs/2508.12223} {arXiv:2508.12223 [cond-mat.soft]} \BibitemShut {NoStop}%
\bibitem [{\citenamefont {Kundu}\ \emph {et~al.}(2025)\citenamefont {Kundu}, \citenamefont {Mondal}, \citenamefont {Biswas}, \citenamefont {Pal},\ and\ \citenamefont {Khan}}]{kundu2025emulating}%
  \BibitemOpen
  \bibfield  {author} {\bibinfo {author} {\bibfnamefont {Sandip}\ \bibnamefont {Kundu}}, \bibinfo {author} {\bibfnamefont {Dibyendu}\ \bibnamefont {Mondal}}, \bibinfo {author} {\bibfnamefont {Arup}\ \bibnamefont {Biswas}}, \bibinfo {author} {\bibfnamefont {Arnab}\ \bibnamefont {Pal}}, \ and\ \bibinfo {author} {\bibfnamefont {Manas}\ \bibnamefont {Khan}},\ }\bibfield  {title} {\enquote {\bibinfo {title} {Emulating microbial run-and-tumble and tactic motion by stochastically reorienting synthetic active Brownian particles},}\ }\href@noop {} {\bibfield  {journal} {\bibinfo  {journal} {arXiv preprint arXiv:2509.21903}\ } (\bibinfo {year} {2025})}\BibitemShut {NoStop}%
\bibitem [{\citenamefont {Bhat}\ \emph {et~al.}(2016)\citenamefont {Bhat}, \citenamefont {De~Bacco},\ and\ \citenamefont {Redner}}]{bhat2016stochastic}%
  \BibitemOpen
  \bibfield  {author} {\bibinfo {author} {\bibfnamefont {Uttam}\ \bibnamefont {Bhat}}, \bibinfo {author} {\bibfnamefont {Caterina}\ \bibnamefont {De~Bacco}}, \ and\ \bibinfo {author} {\bibfnamefont {S}~\bibnamefont {Redner}},\ }\bibfield  {title} {\enquote {\bibinfo {title} {Stochastic search with poisson and deterministic resetting},}\ }\href@noop {} {\bibfield  {journal} {\bibinfo  {journal} {Journal of Statistical Mechanics: Theory and Experiment}\ }\textbf {\bibinfo {volume} {2016}},\ \bibinfo {pages} {083401} (\bibinfo {year} {2016})}\BibitemShut {NoStop}%
\bibitem [{\citenamefont {Evans}\ and\ \citenamefont {Ray}(2025)}]{evans2025stochastic}%
  \BibitemOpen
  \bibfield  {author} {\bibinfo {author} {\bibfnamefont {Martin~R}\ \bibnamefont {Evans}}\ and\ \bibinfo {author} {\bibfnamefont {Somrita}\ \bibnamefont {Ray}},\ }\bibfield  {title} {\enquote {\bibinfo {title} {Stochastic resetting prevails over sharp restart for broad target distributions},}\ }\href@noop {} {\bibfield  {journal} {\bibinfo  {journal} {Physical Review Letters}\ }\textbf {\bibinfo {volume} {134}},\ \bibinfo {pages} {247102} (\bibinfo {year} {2025})}\BibitemShut {NoStop}%
\bibitem [{{\relax DLMF}()}]{NIST:DLMF}%
  \BibitemOpen
  {\relax DLMF},\ \href {https://dlmf.nist.gov/} {\enquote {\bibinfo {title} {{\it NIST Digital Library of Mathematical Functions}},}\ }\bibinfo {howpublished} {\url{https://dlmf.nist.gov/}, Release 1.2.0 of 2024-03-15},\ \bibinfo {note} {f.~W.~J. Olver, A.~B. {Olde Daalhuis}, D.~W. Lozier, B.~I. Schneider, R.~F. Boisvert, C.~W. Clark, B.~R. Miller, B.~V. Saunders, H.~S. Cohl, and M.~A. McClain, eds.}\BibitemShut {Stop}%
\bibitem [{\citenamefont {Nguyen}\ \emph {et~al.}(2024)\citenamefont {Nguyen}, \citenamefont {Pham}, \citenamefont {Ngo}, \citenamefont {Do}, \citenamefont {Li},\ and\ \citenamefont {Phan}}]{nguyen2024remark}%
  \BibitemOpen
  \bibfield  {author} {\bibinfo {author} {\bibfnamefont {Minh~DN}\ \bibnamefont {Nguyen}}, \bibinfo {author} {\bibfnamefont {Phuc~H}\ \bibnamefont {Pham}}, \bibinfo {author} {\bibfnamefont {Khang~V}\ \bibnamefont {Ngo}}, \bibinfo {author} {\bibfnamefont {Van~H}\ \bibnamefont {Do}}, \bibinfo {author} {\bibfnamefont {Shengkai}\ \bibnamefont {Li}}, \ and\ \bibinfo {author} {\bibfnamefont {Trung~V}\ \bibnamefont {Phan}},\ }\bibfield  {title} {\enquote {\bibinfo {title} {Remark on the entropy production of adaptive run-and-tumble chemotaxis},}\ }\href@noop {} {\bibfield  {journal} {\bibinfo  {journal} {Physica A: Statistical Mechanics and its Applications}\ }\textbf {\bibinfo {volume} {634}},\ \bibinfo {pages} {129452} (\bibinfo {year} {2024})}\BibitemShut {NoStop}%
\bibitem [{\citenamefont {Sartori}\ \emph {et~al.}(2014)\citenamefont {Sartori}, \citenamefont {Granger}, \citenamefont {Lee},\ and\ \citenamefont {Horowitz}}]{sartori2014thermodynamic}%
  \BibitemOpen
  \bibfield  {author} {\bibinfo {author} {\bibfnamefont {Pablo}\ \bibnamefont {Sartori}}, \bibinfo {author} {\bibfnamefont {L{\'e}o}\ \bibnamefont {Granger}}, \bibinfo {author} {\bibfnamefont {Chiu~Fan}\ \bibnamefont {Lee}}, \ and\ \bibinfo {author} {\bibfnamefont {Jordan~M}\ \bibnamefont {Horowitz}},\ }\bibfield  {title} {\enquote {\bibinfo {title} {Thermodynamic costs of information processing in sensory adaptation},}\ }\href@noop {} {\bibfield  {journal} {\bibinfo  {journal} {PLoS Computational Biology}\ }\textbf {\bibinfo {volume} {10}},\ \bibinfo {pages} {e1003974} (\bibinfo {year} {2014})}\BibitemShut {NoStop}%
\bibitem [{\citenamefont {Lan}\ \emph {et~al.}(2012)\citenamefont {Lan}, \citenamefont {Sartori}, \citenamefont {Neumann}, \citenamefont {Sourjik},\ and\ \citenamefont {Tu}}]{lan2012energy}%
  \BibitemOpen
  \bibfield  {author} {\bibinfo {author} {\bibfnamefont {Ganhui}\ \bibnamefont {Lan}}, \bibinfo {author} {\bibfnamefont {Pablo}\ \bibnamefont {Sartori}}, \bibinfo {author} {\bibfnamefont {Silke}\ \bibnamefont {Neumann}}, \bibinfo {author} {\bibfnamefont {Victor}\ \bibnamefont {Sourjik}}, \ and\ \bibinfo {author} {\bibfnamefont {Yuhai}\ \bibnamefont {Tu}},\ }\bibfield  {title} {\enquote {\bibinfo {title} {The energy--speed--accuracy trade-off in sensory adaptation},}\ }\href@noop {} {\bibfield  {journal} {\bibinfo  {journal} {Nature Physics}\ }\textbf {\bibinfo {volume} {8}},\ \bibinfo {pages} {422--428} (\bibinfo {year} {2012})}\BibitemShut {NoStop}%
\bibitem [{\citenamefont {Welker}\ and\ \citenamefont {Pietzonka}(2026)}]{welker2026accuracycomescostoptimal}%
  \BibitemOpen
  \bibfield  {author} {\bibinfo {author} {\bibfnamefont {Till}\ \bibnamefont {Welker}}\ and\ \bibinfo {author} {\bibfnamefont {Patrick}\ \bibnamefont {Pietzonka}},\ }\href {https://arxiv.org/abs/2602.13173} {\enquote {\bibinfo {title} {Accuracy comes at a cost: Optimal localisation against a flow},}\ } (\bibinfo {year} {2026}),\ \Eprint {http://arxiv.org/abs/2602.13173} {arXiv:2602.13173 [cond-mat.stat-mech]} \BibitemShut {NoStop}%
\end{thebibliography}

%

\end{document}